\documentclass[twocolumn,english,prl,showpacs,preprintnumbers,superscriptaddress]{revtex4-1}
\usepackage[utf8]{inputenc}
\usepackage{color}
\usepackage{amsmath}
\usepackage{amssymb}
\usepackage{graphicx}
\usepackage{textcomp}
\usepackage{dcolumn}
\usepackage{bm}
\usepackage[right]{eurosym}
\usepackage{float}
\usepackage[english]{babel}
\usepackage{blindtext}
\usepackage{babel}
\usepackage[normalem]{ulem}
\DeclareMathOperator\erfc{erfc}

\begin{document}

\title{Time-resolved study of resonant interatomic Coulombic decay in helium nanodroplets}

\author{A. C. LaForge}
\email{aaron.laforge@uconn.edu}
\affiliation{Department of Physics, University of Connecticut, Storrs, Connecticut, 06269, USA}
\affiliation{Physikalisches Institut, Universit{\"a}t Freiburg, 79104 Freiburg, Germany}
\author{R. Michiels}
\affiliation{Physikalisches Institut, Universit{\"a}t Freiburg, 79104 Freiburg, Germany}
\author{Y. Ovcharenko} 
\affiliation{Institut f\"ur Optik und Atomare Physik, Technische Universit\"at, 10623 Berlin, Germany}
\author{A. Ngai}
\affiliation{Physikalisches Institut, Universit{\"a}t Freiburg, 79104 Freiburg, Germany}
\author{J. M. Escart\'{\i}n}
\affiliation{Institut de Qu\'{\i}mica Te\`{o}rica i Computacional, Universitat de Barcelona, Carrer de Mart\'{\i} i Franqu\`{e}s 1,08028 Barcelona, Spain.}
\author{N. Berrah}
\affiliation{Department of Physics, University of Connecticut, Storrs, Connecticut, 06269, USA}
\author{C. Callegari}
\affiliation{Elettra-Sincrotrone Trieste, 34149 Basovizza, Trieste, Italy}
\author{A. Clark} 
\affiliation{Laboratoire Chimie Physique Mol{\'e}culaire, Ecole Polytechnique F{\'e}d{\'e}rale de Lausanne, 1015 Lausanne, Switzerland}
\author{M. Coreno} 
\affiliation{CNR - Istituto di Struttura della Materia, 00016 Monterotondo Scalo, Italy}
\author{R. Cucini}
\affiliation{Elettra-Sincrotrone Trieste, 34149 Basovizza, Trieste, Italy}
\author{M. Di Fraia} 
\affiliation{Elettra-Sincrotrone Trieste, 34149 Basovizza, Trieste, Italy}
\author{M. Drabbels} 
\affiliation{Laboratoire Chimie Physique Mol{\'e}culaire, Ecole Polytechnique F{\'e}d{\'e}rale de Lausanne, 1015 Lausanne, Switzerland}
\author{E. Fasshauer} 
\affiliation{Department of Physics and Astronomy, Aarhus University, 8000 Aarhus C, Denmark}
\author{P. Finetti}
\affiliation{Elettra-Sincrotrone Trieste, 34149 Basovizza, Trieste, Italy}
\author{L. Giannessi}
\affiliation{Elettra-Sincrotrone Trieste, 34149 Basovizza, Trieste, Italy}
\affiliation{Nazionale di Fisica Nucleare - Laboratori Nazionali di Frascati,  Via E. Fermi 40, 00044 Frascati, Roma, Italy}
\author{C. Grazioli} 
\affiliation{CNR - Istituto di Struttura della Materia, 00016 Monterotondo Scalo, Italy}
\author{D. Iablonskyi} 
\affiliation{Institute of Multidisciplinary Research for Advanced Materials, Tohoku University, Sendai  980-8577, Japan}
\author{B. Langbehn} 
\affiliation{Institut f\"ur Optik und Atomare Physik, Technische Universit\"at, 10623 Berlin, Germany}
\author{T. Nishiyama} 
\affiliation{Division of Physics and Astronomy, Graduate School of Science, Kyoto  University,  Kyoto  606-8502, Japan}
\author{V. Oliver} 
\affiliation{Laboratoire Chimie Physique Mol{\'e}culaire, Ecole Polytechnique F{\'e}d{\'e}rale de Lausanne, 1015 Lausanne, Switzerland}
\author{P. Piseri}
\affiliation{Dipartimento di Fisica and CIMaINa, Università degli Studi di Milano, 20133 Milano, Italy}
\author{O. Plekan}
\affiliation{Elettra-Sincrotrone Trieste, 34149 Basovizza, Trieste, Italy}
\author{K. C. Prince}
\affiliation{Elettra-Sincrotrone Trieste, 34149 Basovizza, Trieste, Italy}
\author{D. Rupp} 
\affiliation{Institut f\"ur Optik und Atomare Physik, Technische Universit\"at, 10623 Berlin, Germany}
\author{S. Stranges} 
\affiliation{Department of Chemistry, University Sapienza, 00185 Rome, Italy}
\author{K. Ueda} 
\affiliation{Institute of Multidisciplinary Research for Advanced Materials, Tohoku University, Sendai 980-8577, Japan}
\author{N. Sisourat}
\affiliation{Sorbonne Universit\'e, CNRS, Laboratoire de Chimie Physique Mati\`ere et Rayonnement, UMR 7614, F-75005 Paris, France}
\author{J. Eloranta}
\affiliation{Department of Chemistry and Biochemistry, California State University at Northridge, Northridge, CA, 91330, USA} 
\author{M. Pi}
\affiliation{Departament FQA, Facultat de F\'{\i}sica, Universitat de Barcelona, Barcelona, 08028, Spain}
\affiliation{Institute of Nanoscience and Nanotechnology (IN2UB), Universitat de Barcelona, Barcelona, 08028, Spain}	
\author{M. Barranco}
\affiliation{Departament FQA, Facultat de F\'{\i}sica, Universitat de Barcelona, Barcelona, 08028, Spain}
\affiliation{Institute of Nanoscience and Nanotechnology (IN2UB), Universitat de Barcelona, Barcelona, 08028, Spain}	
\author{F. Stienkemeier}
\affiliation{Physikalisches Institut, Universit{\"a}t Freiburg, 79104 Freiburg, Germany}
\author{T. M\"oller} 
\email{thomas.moeller@physik.tu-berlin.de}
\affiliation{Institut f\"ur Optik und Atomare Physik, Technische Universit\"at, 10623 Berlin, Germany}
\author{M. Mudrich} 
\email{mudrich@phys.au.dk}
\affiliation{Department of Physics and Astronomy, Aarhus University, 8000 Aarhus C, Denmark}

\begin{abstract}
	
When weakly-bound complexes are multiply excited by intense electromagnetic radiation, energy can be exchanged between neighboring atoms through a type of resonant interatomic Coulombic decay (ICD). This decay mechanism due to multiple excitations has been predicted to be relatively slow, typically lasting tens to hundreds of picoseconds. Here, we directly measure the ICD timescale in resonantly excited helium droplets using a high resolution, tunable, extreme ultraviolet free-electron laser. Over an extensive range of droplet sizes and laser intensities, we discover the decay to be surprisingly fast, with decay times as fast as 400 femtoseconds, and to only present a weak dependence on the density of the excited states. Using a combination of time dependent density functional theory and \textit{ab initio} quantum chemistry calculations, we elucidate the mechanisms of this ultrafast decay process where pairs of excited helium atoms in one droplet strongly attract each other and form merging void bubbles which drastically accelerates ICD.

\end{abstract}

\date{\today}

\maketitle


\begin{figure}
	\begin{center}{
			\includegraphics[width=\linewidth]{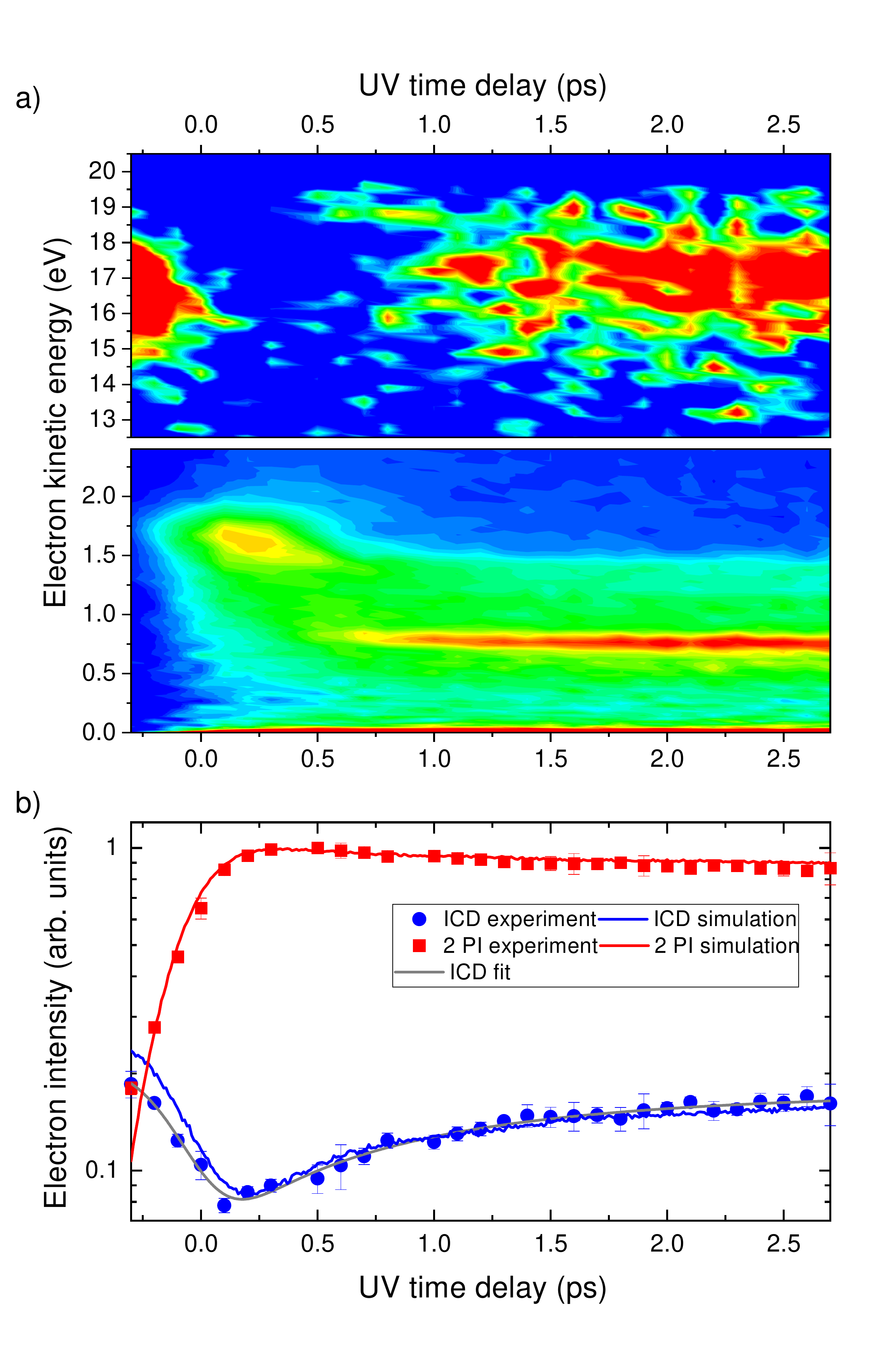}}
		\caption{a) Time-resolved electron kinetic energy distributions of resonantly excited He droplets centered around the ICD peak (top panel) and the two-photon ionization (2PI) signal (bottom panel). b) Projection of the intensity of the ICD peak (blue circles) and the 2PI signal (red squares) as a function of XUV-UV pump-probe delay. The experimental data is fitted with a convoluted exponential decay function (gray line). The red and blue lines show the results of a MC simulation (see text for details). The droplet size was 76,000 atoms and the excitation photon energy was 21.6\,eV.}
		\label{fig1}
	\end{center}
\end{figure}

\section*{Introduction}

Short-wavelength free-electron lasers (FELs) are well-suited for studying light-matter interactions, due to their high intensity and ultrashort pulse duration, where many photons can be absorbed in a system within a few femtoseconds. For condensed systems, the complexity of interatomic processes makes gaining a thorough understanding of the ionization mechanisms and dynamics tedious, if not impossible. On the other hand, free, weakly-bound nanosystems such as van der Waals (vdW) clusters, can be used to study such interatomic interactions in a well-controlled manner. In particular, the study of vdW clusters irradiated by intense FEL radiation has led to the observation of numerous interatomic processes\,\cite{Wabnitz2002,Bostedt2008,Bostedt2010}.

A novel type of interatomic process known as interatomic Coulombic decay (ICD)~\cite{Cederbaum1997} has been widely studied in weakly-bound systems~\cite{Hergenhahn2011,Jahnke2015}. In cases where local Auger decay is energetically forbidden, an excited atom or molecule releases its excitation energy by transferring it to a neighboring atom or molecule, which can result in its ionization. In general, ICD is a prominent decay mechanism in a multitude of systems, specifically those of biological relevance~\cite{Gokhberg2014,Trinter2014,Stumpf2016,Ren2018}. One of the factors determining the importance of ICD in a nanosystem is its decay time, which is directly linked to its efficiency. With the advent of seeded FELs and the availability of intense, tunable extreme ultra-violet (XUV) radiation~\cite{Allaria2012,Allaria2012a}, new types of resonant ICD~\cite{Kuleff2010} have been observed in vdW clusters~\cite{LaForge2014,Ovcharenko2014,Iablonskyi2016,Ovcharenko2020} where energy is exchanged between neighboring excited atoms. Additionally, similar resonant-type ICDs were observed by synchrotron radiation in mixed vdW clusters where energy was exchanged between species~\cite{Trinter2013,Buchta2013,LaForge2019}.  

He nanodroplets have served as model systems for studying interatomic processes induced by one photon~\cite{Havermeier2010,Sisourat2010,Buchta2013,Stumpf2014,LaForge2016,Shcherbinin2017,LaForge2019,BenLtaief2019,BenLtaief2020} and multiple photons~\cite{LaForge2014,Ovcharenko2014,Katzy2015,Ovcharenko2020}, due to their simple electronic structure and extremely weak atom-atom interactions. Moreover, beyond being a testbench for studying atomic and binary interatomic processes, He nanodroplets are quantum fluid clusters, which exhibit unique features such as voids, or ``bubbles'', around impurities~\cite{Haeften2002,Thaler2018,Mudrich2020}, which can freely move about the droplet owing to its superfluid state~\cite{Toennies2004}. 

Here, we report on time-resolved measurements of resonant ICD in He droplets. The process is initiated by an XUV pulse tuned to the resonant 1s2p droplet band ($h\nu$\,=\,21.6\,eV)~\cite{Joppien1993}. This creates multiple excited atoms in the droplet which can decay via ICD, \textit{i.~e.} the energy from one excited He atom, He$^*$, is transferred to another He$^*$, which is then ionized. A second, time-delayed UV pulse can directly ionize the excited atom(s) in the droplet thereby interrupting and halting any interatomic decay processes. Over an extensive range of droplet sizes and laser pulse energies, the decay mechanism was found to be much faster than predicted by theory~\cite{Kuleff2010}. Even more surprising, the decay rate is nearly independent of the number of excited atoms per droplet, although theory predicts a very strong dependence on the internuclear distance. To understand the discrepancies, the experimental results were modelled using a combination of time-dependent density-functional theory (TDDFT)~\cite{Ancilotto:2017} and \textit{ab initio} calculations of the doubly excited He$^*$-He$^*$ pair potentials as well as the ICD widths. We discovered that the ICD dynamics are largely determined by the attractive interaction of closely-spaced He$^*$ atoms and by the formation of bubbles around them. The latter strongly accelerates the ICD via the merging of overlapping bubbles. The results show that interatomic processes in condensed phase nanosystems are governed by a complex set of relaxation mechanisms which can result in ultrafast autoionization.

\section*{Experiment}

This work was performed at the Low Density Matter endstation~\cite{Lyamayev2013} of the seeded FEL FERMI, in Trieste, Italy. The FEL photon energy ($21.6$\,eV) was tuned via the seed laser, undulator gaps, and other machine parameters, yielding a pulse length of approximately 100\,fs full width at half maximum (FWHM)~\cite{Allaria2012,Allaria2012a}. The measurements were taken over two separate periods with different experimental conditions. The second set of parameters is given in parentheses.  The FEL pulse energy, varied from 0.1\,$\mu$J to 50\,$\mu$J, was determined upstream by gas ionization, taking the nominal reflectivity of the optical elements in the beam transport system into account. The diameter of the FEL focus was 250\,$\mu$m FWHM. The UV probe pulse was obtained from a frequency-tripled (-doubled) Ti:Sapphire laser ($h\nu'$\,=\,4.8 (3.2)\,eV) with a pulse energy of 50 (200)\,$\mu$J with a focus diameter  of 250\,$\mu$m FWHM. A tin filter of 160\,$\mu$m thickness was used to suppress higher order harmonic radiation. The cross correlation between the FEL and the probe laser was 200\,fs FWHM, measured by resonant two-photon ionization of He. A supersonic gas jet of He nanodroplets was produced by expansion of high pressure He gas through a pulsed, cryogenically cooled Even-Lavie nozzle. By varying the expansion conditions (backing pressure and nozzle temperature), the mean cluster size was varied in the range of $\langle N\rangle =10^2$-$10^5$ He atoms. The nanodroplet beam was perpendicularly crossed by the FEL and UV beams at the center of a velocity map imaging spectrometer~\cite{Lyamayev2013}. The electron kinetic energy distributions were reconstructed using the Maximum Entropy Legendre Reconstruction method~\cite{Dick2014}.


\section*{Results and Discussion}

Fig.~\ref{fig1}\,a) shows the distributions of electron kinetic energy $E_e$ emitted by resonantly excited He droplets as a function of the delay between XUV pump and UV probe laser pulses. The mean droplet size was $\langle N\rangle = 76,000$ atoms and the XUV intensity was $2.8\times 10^9$\,W/cm$^{2}$. At low kinetic energies ($0 < E_e < 2$\,eV), the electron distribution is created by resonant two-photon ionization (2 PI) in He nanodroplets. At short delays, $\Delta t < 1$\,ps, this shows the droplet-induced relaxation dynamics of He$^*$ from the XUV-excited 1s2p state to the 1s2s state\,\cite{Mudrich2020}. At higher kinetic energies ($15 < E_e < 18\,$\,eV), resonant multiphoton ICD is observed according to the reaction~\cite{Kuleff2010,LaForge2014,Ovcharenko2014}
\[
(\mathrm{He}^* + \mathrm{He}^*)\mathrm{He}_{N-2} \rightarrow (\mathrm{He}^+ + e_\mathrm{ICD} + \mathrm{He})\mathrm{He}_{N-2}.
\]
Here, He$_N$ denotes the He droplet and $e_\mathrm{ICD}$ is the ICD electron. A discussion of the electronic states which initiate this type of ICD is given in Appendix 1. Since ICD is a binary process, at least two excited atoms are required per droplet. 

The intensities of photoelectrons (red squares) and ICD electrons (blue circles), depicted in Fig.~\ref{fig1}\,b), display opposing trends in their time evolution: the 2 PI signal is enhanced at delays $0 < \Delta t < 0.2$\,ps whereas the ICD signal is depressed. This can be rationalized by the depletion of the He$^*$ population through photoionization by the UV probe pulse, thereby suppressing the ICD. As the pump-probe delay is increased, ICD can proceed before the He$^*$ are photoionized and the $e_\mathrm{ICD}$ yield is replenished. Thus, the rise of the $e_\mathrm{ICD}$ yield reflects the timescale of the ICD. To quantify this process, the ICD signal was fitted with a function [gray line in  Fig.\,\ref{fig1}\,b)] accounting for the exponential rise as well as the temporal overlap of two Gaussian pulses near time zero. A thorough discussion of this fitting procedure is given in Appendix 2. Fig.~\ref{fig1}\,b) also shows the ICD (blue line) and 2 PI (red line) data from a Monte Carlo (MC) simulation as discussed later in the text. Additional data and fits for different experimental parameters are given in Appendix 3. 

\begin{figure}
	\begin{center}{
			\includegraphics[width=\linewidth]{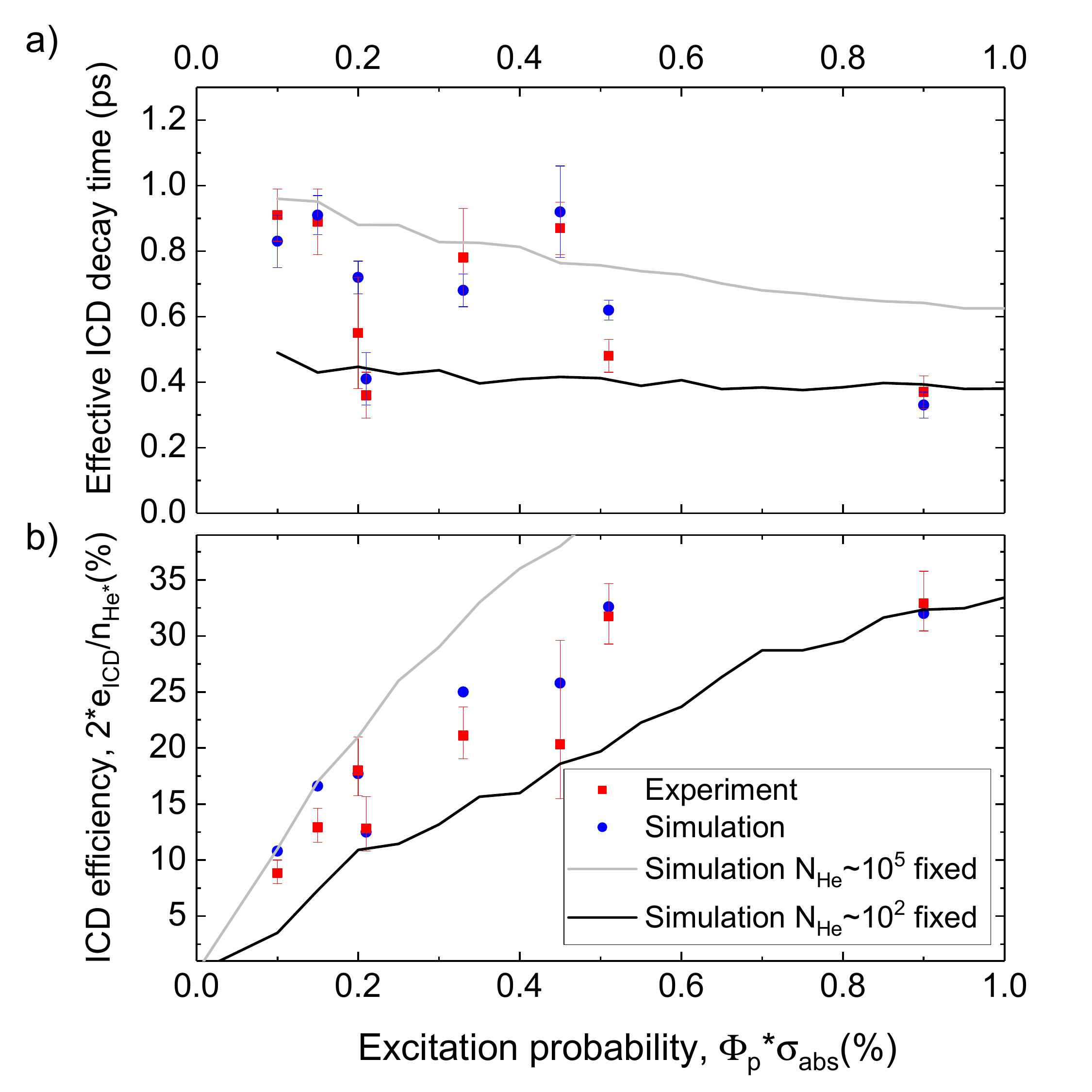}}
		\caption{a) Effective ICD decay time and b) ICD efficiency plotted as a function of the excitation probability (red squares). MC simulation results for the specific experimental conditions, droplet size and FEL intensity, are shown as blue dots. Additionally, to show the general trend in the simulations on the FEL intensity, the results for fixed small and large droplets are shown as black and gray lines, respectively.}
		\label{fig2}
	\end{center}
\end{figure}

To systematically investigate the dynamics of ICD in He nanodroplets, pump-probe delay dependencies over a wide range of He droplet sizes and XUV intensities were recorded. The latter controls the He$^*$ excitation probability (photon flux$\,\times\,$absorption cross section) and thereby the mean distance between He$^*$ in a droplet. Due to the strong coupling between the FEL power, droplet size, and collective auto-ionization (CAI) effects~\cite{Ovcharenko2020}, only a limited range of excitation probabilities (0.1-1\,\%) showed a clearly distinguishable ICD peak, despite the broad range of droplet sizes and FEL intensities available. As the FEL intensity increases, multiple excited atoms may interact, leading to decay by CAI and formation of a nanoplasma~\cite{Ovcharenko2014}. In the transition from ICD to CAI, the ICD peak broadens and shifts to lower energies due to the formation of a collective Coulomb potential, and eventually becomes dominated by low-energy thermal electrons from the nanoplasma~\cite{Ovcharenko2020}.

Similar to what is shown in Fig.~\ref{fig1}\,b), each ICD delay dependence is fitted with a function to determine the time constant of the $e_\mathrm{ICD}$ evolution. The resulting ICD times, $\tau_\mathrm{ICD}$, and $e_\mathrm{ICD}$ yields are respectively plotted as red symbols in Fig.~\ref{fig2}\,a) and b) as a function of the excitation probability. The corresponding MC simulation results are shown as blue dots. The $e_\mathrm{ICD}$ yield is determined from the total number of detected electrons and the He$^*$ photoionization cross section~\cite{chang1995effect}. It is normalized to the number of He$^*$ atoms in the droplet and multiplied by two to account for the fact that two excitations produce one ICD electron. The resulting ICD efficiency rises from 0.09 to 0.32 in the given range of He$^*$ excitation probability, while $\tau_\mathrm{ICD}$ decreases from 1000 to 400\,fs. To decouple the effect of the droplet size from the FEL intensity, we have additionally performed MC simulations (see the SM for details) for fixed droplet sizes. The results for small and large droplets are shown in Fig.~\ref{fig2} as black and gray lines, respectively. For small droplets, the ICD time is nearly constant and lower than the ICD decay times for large droplets, which show a weak dependence on the FEL intensity. The ICD efficiency shown in Fig.~\ref{fig2}\,b) rises from zero as a function of the excitation probability with a higher slope for large droplets. Overall, the MC simulations are in excellent agreement with the experimental data.

\begin{figure*}
	\begin{center}{
			\includegraphics[width=1\textwidth]{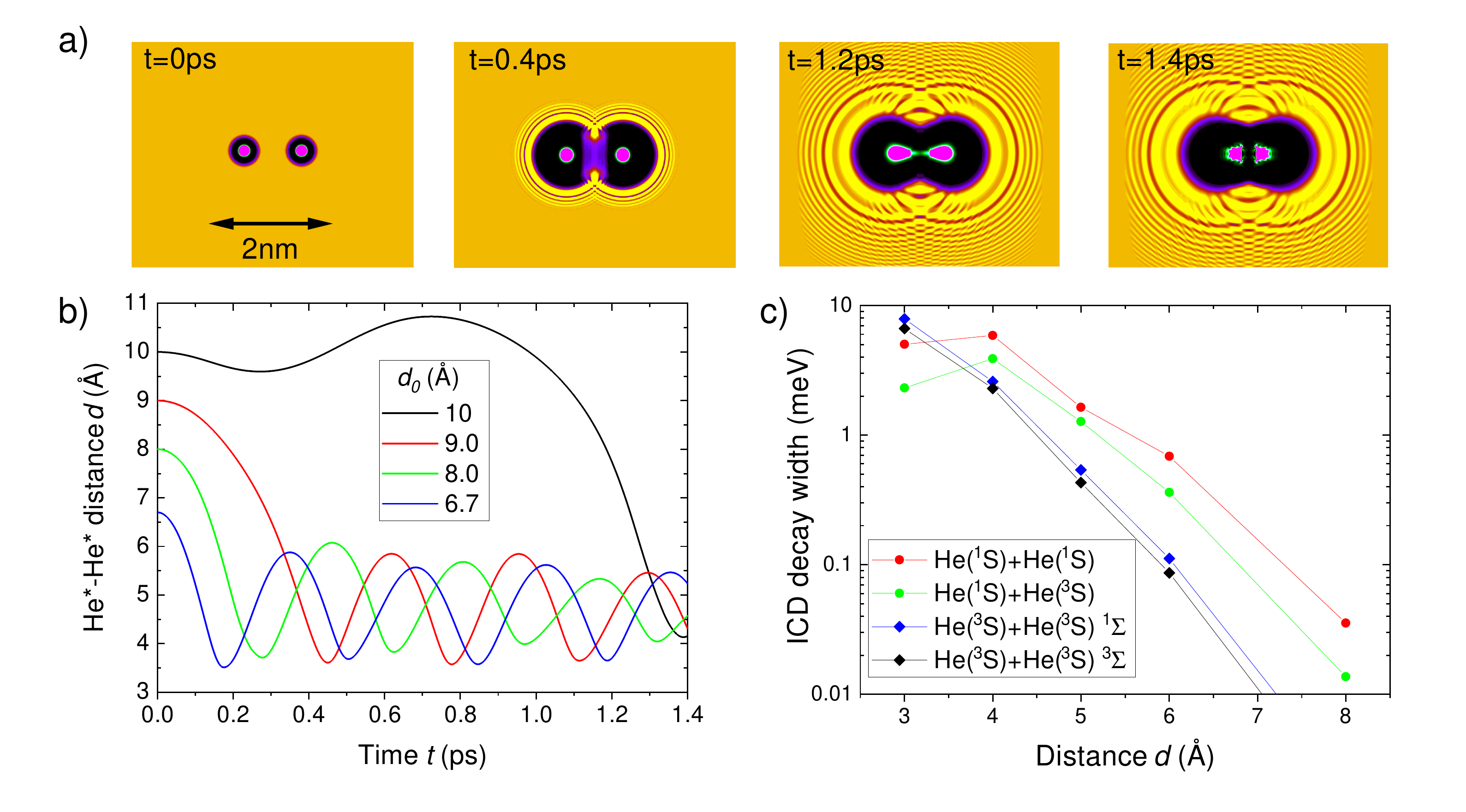}}
		\caption{a) Snapshots of the He density evolution around two He$^*$ centers separated initially by $d_0=10$\,\AA. The probability distribution of He$^*$ is represented as pink dots and the black areas show the void bubbles forming. b) Evolution  of the He$^*$-He$^*$ distance $d$ for various values of the initial distance $d_0$. c) Calculated ICD widths for various combinations of 1s2s-excited He$^*$ atomic states.}
		\label{figX}
	\end{center}
\end{figure*}

In general, the measured ICD decay times are surprisingly short ($\tau_\mathrm{ICD}<1\,$ps) compared to estimates based on the virtual photon ICD model for this type of system, which yield 52\,ps for the fastest channel~\cite{Kuleff2010}. Furthermore, previous static measurements predicted this type of ICD to be much slower, in the high ps range~\cite{Iablonskyi2016,BenLtaief2019}. Further proof of the discrepancy between theory and experimental results can be seen in Fig.~\ref{figX}\,c), which shows the ICD decay width, $\Gamma$, as a function of the He$^*$-He$^*$ distance. $\Gamma (d)$ is calculated by the Fano-CI-Stieltjes method~\cite{miteva2017computations} for all possible combinations of electronic states populated during droplet relaxation~\cite{Mudrich2020,BenLtaief2019} (see the SM for details). $\Gamma$, which is inversely proportional to the decay time, $\tau_\mathrm{ICD}$, shows a very strong dependence on the He$^*$-He$^*$ distance. On the other hand, the measured $\tau_\mathrm{ICD}$, in Fig.~\ref{fig2}\,a), shows only a weak dependence on the He$^*$ excitation probability, which is a measurable quantity proportional to the mean He$^*$-He$^*$ distance. The observed ultrafast ICD rates, in the fs regime, can only be explained through an additional mechanism that brings the two He$^*$ atoms into close contact. Excitation migration~\cite{Scharf1986,scheidemann1993anomalies}, excitation delocalization~\cite{closser2014simulations} and hole hopping have been discussed extensively over the years, especially in the context of Penning ionization~\cite{scheidemann1993anomalies,seong1998short}. While fast excitation transfer, akin to exciton hopping, can explain the high efficiency of the Penning process~\cite{scheidemann1993anomalies}, it cannot account for the short ICD lifetime. Delocalization of excitations over an extended region of the He droplet as a consequence of exciton hopping would lead to a reduced local spatial overlap and thus to low ICD rates. Besides, the large variation of the interatomic distances between He atoms in the droplets due to the large zero-point motion as well as many-body quantum effects may also limit delocalization~\cite{seong1998short,closser2014simulations}. Unfortunately, the problem of excitation transfer in superfluid He has not yet been addressed theoretically, despite the numerous experimental Penning ionization studies. That said, an additional mechanism is required that brings two He$^*$ in close contact such that ICD takes place at short distances.

Aside from the fast delay-time dependence of the ICD signal, we observed that the $e_\mathrm{ICD}$ yield in most cases does not fully rise to the level measured at negative delays within the full range of pump-probe delays, see Appendix 2 and 3. This indicates that some of the He$^*$ decay by ICD much more slowly than the experimentally observed convergence from which we deduce $\tau_\mathrm{ICD}$. Furthermore, the observation that the ICD efficiency never exceeds 35\% in our experiments points at a competing relaxation channel that prevents the majority of He$^*$ from decaying via ICD. 

To better understand the response of He nanodroplets to multiple excitations and to rationalize our experimental findings, time-dependent density functional (TDDFT) simulations were performed~\cite{Barranco:2006,Ancilotto:2017,Mudrich2020,Hernando:2012} for the motion of He$^*$ pairs. To keep the simulations tractable, we considered bulk superfluid He, which is coupled to the He$^*$ pair self-consistently. Due to the light mass of the He$^*$ “impurities”, they must be treated quantum mechanically with the potential term given by the He$^*$-droplet interaction. To include the interaction between the two He$^*$ atoms, the He$^*$-He$^*$ pair potentials were calculated using highly correlated\,\textit{ab initio} methods (see the SM for details). 

Fig.~\ref{figX}\,a) shows the time evolution of the 2D cuts of the He density distribution (yellow-red area) when the two excited He atoms were initially separated by $d_0=10$\,\AA\,(pink-green dots). Animation of these simulated dynamics for various initial conditions are included in the SM. Upon excitation, bubbles form around them due to the repulsion between the Rydberg electrons and the
surrounding closed shell He atoms~\cite{Buchenau1991,Hernando:2012,Haeften2002,Thaler2018,Mudrich2020}. As the bubbles grow and the two He$^*$ atoms weakly attract each other, the bubbles eventually overlap and merge into one large bubble. The salient feature is that shortly after the two bubbles coalesce, the two He$^*$ are strongly accelerated towards each other. This process is facilitated by the merging of the bubbles where the He$^*$s reach interatomic distances $d<4$\,\AA\,within 400\,fs for all initial distances $d_0$ up to 9\,\AA, see Fig.\,\ref{figX}\,(b). As ICD is not explicitly included in the TDDFT simulations, the He$^*$ pair continues vibrating at short distance due to the attractive He$^*$-He$^*$ potential. However, within the first half cycle of the vibration, the ICD decay width reaches $\Gamma (d=4\,\mathrm{\AA})=5.9$\,meV, corresponding to a characteristic ICD time $\tau_\mathrm{ICD}^{theo}=110$\,fs for the He$^*$($^1$S)+He$^*$($^1$S) pair, which has the largest branching ratio in the droplet relaxation\,\cite{BenLtaief2019,Mudrich2020}. Thus, all He$^*$ pairs with $d_0\lesssim 10\,$\AA\,actually decay via ICD within $t\lesssim 1.4\,$ps with a probability of near unity. Thus, we conclude that the decay in this particular system is largely determined by the pair-wise attraction of excited atoms, as well as the quantum fluid dynamics of the merging bubbles. 

For larger initial distances ($d_0>10\,$\AA), the time between He$^*$ excitation and bubble merging quickly increases to $t>10\,$ps and therefore ICD becomes very slow. This explains the observed incomplete replenishment of the $e_\mathrm{ICD}$-signal at long pump-probe delays. But why do not all He$^*$ decay by ICD in the absence of the probe pulse? It is known that radiative decay from He clusters is not expected to play a significant role  since the lifetime is in the ns regime\,\cite{Haeften2002,Haeften2011}. Previous experimental and theoretical studies have shown that, following the bubble formation, some of the He$^*$ remain weakly bound to the He droplet surface where they eventually form He$_2^*$ excimers\,\cite{nijjar2018conversion}, whereas others are directly ejected from the droplets\,\cite{Buchenau1991,moller1999photochemistry,Mudrich2020}. Once a He$^*$ has detached from the droplet, it can no longer decay via ICD, but it still contributes to the photoionization signal. Based on our measurements (Fig.\,\ref{fig2}\,b)), the fraction of ejected He$^*$ was estimated to be larger than 50\%. The competition between direct ejection and ICD for initial distances $d_0>10\,$\AA\, is strongly dependent on the droplet size.   

The ICD dynamics in He nanodroplets are largely governed by the motion of He$^*$ driven by the bubble dynamics and the interatomic He$^*$-He$^*$ potential, competing against the ejection of surface He$^*$s from the droplet. To account for the aforementioned effects, a simplified MC simulation based on $\Gamma (d)$ was developed, the results are displayed in Fig.\,\ref{fig2}\,b). The He droplet was treated as homogeneously-packed He atoms represented by same-sized spheres. An initial number of He$^*$s, according to the XUV intensity and the He droplet absorption cross section\,\cite{Joppien1993}, were placed at random positions within the droplet. Then, for each He$^*$ the following conditions were tested. If the distance to the droplet surface $d_S<7.5$\,\AA\,and the distance to the nearest neighbor He$^*$ $d_0>9.5$\,\AA, then the He$^*$ is ejected. If $d_S>7.5$\,\AA\,and $d_0<15.5$\,\AA, then the He$^*$ undergoes ICD; the ICD probability was then calculated based on $\Gamma (d)$ according to the trajectory $d(t)$ obtained from the TDDFT simulations. If $d_S>7.5$\,\AA\,and $d_0>15.5$\,\AA, the He$^*$ will not decay by ICD and only photoionization is possible. The probe pulse was implemented by converting the He$^*$ into photoelectrons at a rate consistent with the experimental estimate. The values $d_0=15.5$\,\AA~and $d_S=7.5$\,\AA, used as criteria for ICD enhanced by bubble merging and He$^*$ ejection, respectively, were deduced from the TDDFT simulations. Additionally, when the same simulation was performed for fixed positions of He$^*$, the ICD time constants were 1-2 orders of magnitude longer than the experimental values, thus demonstrating the importance of ultrashort bubble dynamics and the attractive He$^*$-He$^*$ potential. An in-depth discussion of these simulations is given in Appendix 4. 

\section*{Conclusions}

To summarize, we have performed time-resolved measurements of resonant ICD in He nanodroplets. Over a wide range of droplet sizes and laser pulse energies, we have found the decay to be as fast as 400\,fs, and to have little dependence on the density of excited states, in contrast to the strong dependence of the predicted ICD decay width on the distance between excitations. Our simulations have shown that the ICD dynamics is largely determined by the pair-wise attraction of excited atoms, as well as the peculiar response of He droplets to multiple resonant excitations. The formation of bubbles around the excitations and their subsequent merging accelerates ICD, whereas the ejection of excited state atoms from the droplet competes with it. While excited state bubble dynamics is a phenomenon unique to fluids, our time-resolved results nevertheless have clearly demonstrated that ICD in the condensed phase is governed by complex, ultrafast relaxation mechanisms that can couple translational, electronic, and spin degrees of freedom. In general, He nanodroplets are an ideal platform to study processes relevant to a broad range of fields from the molecular to the condensed phase.  

\section*{Acknowledgements}

The authors gratefully acknowledge financial support from  the Carl-Zeiss-Stiftung, the Deutsche Forschungsgemeinschaft (DFG) under grant MO 719/14-2 and within the frame of the Priority Programme 1840 ‘Quantum Dynamics in Tailored Intense Fields (MU 2347/12-1 and STI 125/22-2), and the Carlsberg Foundation. TDDFT work has been performed under grant   FIS2017-87801-P (AEI/FEDER, UE) (M.B., M.P.). A.C.L. and N.B. acknowledge the support of the Chemical Sciences, Geosciences and Biosciences Division, Office of Basic Energy Sciences, Office of Science, US Department of Energy, grant no DE-SC0012376. J.M.E. acknowledges support from Ministerio de Ciencia e Innovaci\'{o}n of Spain through the Unidades de Excelencia ``Mar\'{\i}a de Maeztu'' grant MDM-2017-0767.

	\begin{figure}
	\includegraphics[width=\linewidth]{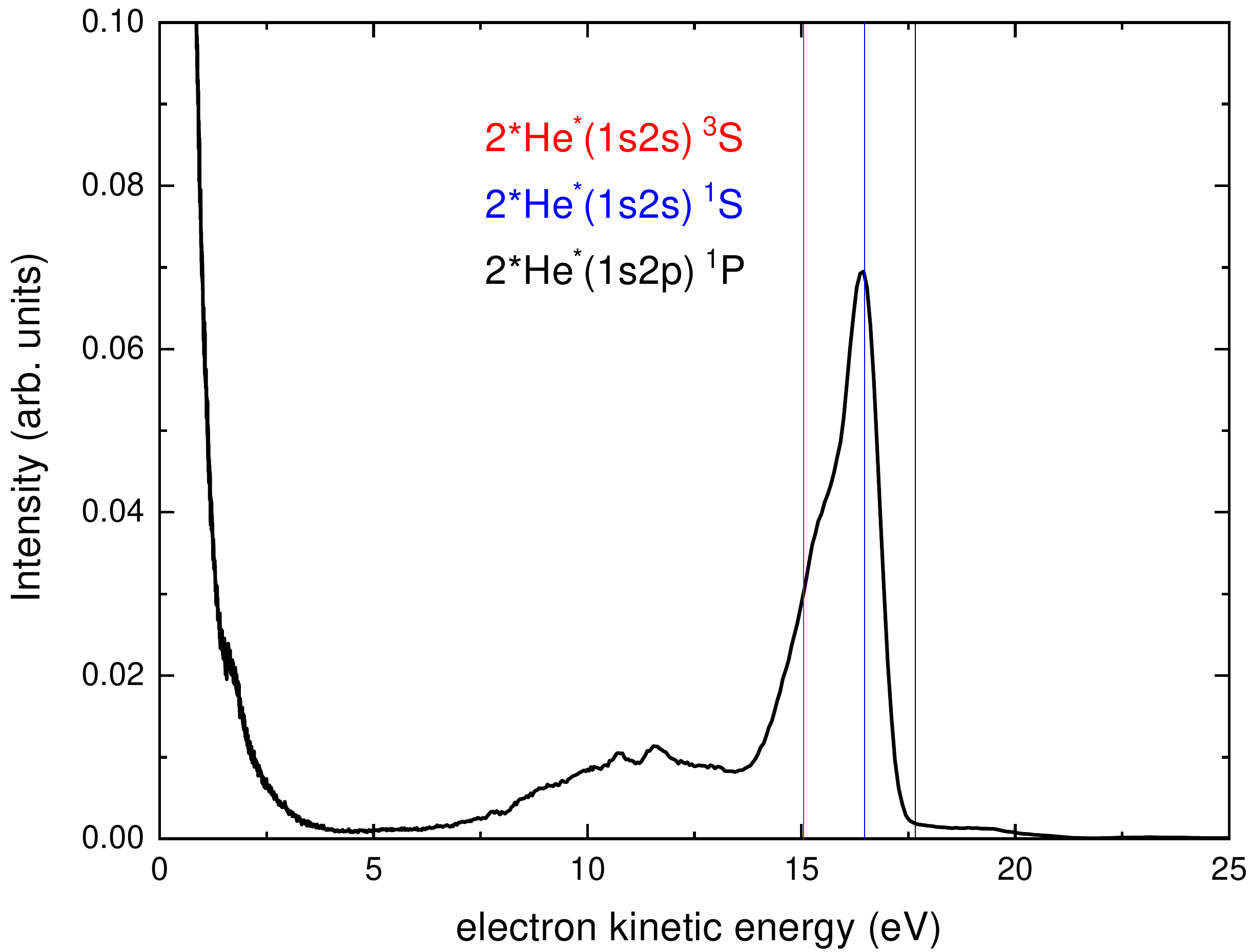}\caption{Static (XUV only) electron kinetic energy distribution measured at $h\nu = 23.7$\,eV. The vertical lines depict the nominal values of ICD electron energies for pairs of He$^*$ atoms in the three lowest excited states.\label{fig:ICD:Static}}
	\end{figure}
	
\section*{Appendix 1: High resolution ICD electron kinetic energy distribution} 
Information about the electronic states involved in the ICD process is encoded in the kinetic energy distribution of ICD electrons. Fig.~\ref{fig:ICD:Static} shows a high-resolution electron spectrum measured at the photon energy $h\nu = 23.7$\,eV. At this photon energy, the 1s4p excited state of He droplets is resonantly excited~\cite{Joppien1993}. The mean droplet size was set to $5\times 10^5$ He atoms. Besides the large signal component at low kinetic energy resulting from CAI~\cite{Ovcharenko2014}, an additional peak is observed around 16\,eV with a shoulder near 15\,eV, which is due to ICD. For comparison, we added vertical lines showing the expected ICD electron energies, $E_{e,\, \mathrm{ICD}}$, for pairs of He$^*$ in the lowest excited states 2s2s\,$^{1,\, 3}S$ and 2s2p\,$^{1}P$ according to
\begin{equation}
E_{e,\, \mathrm{ICD}} = 2E_{\mathrm{He}(1s2s,p)} - E_{i,\,\mathrm{He}}.
\label{eqICD}
\end{equation}
Here $E_{\mathrm{He}(\text{1s2s,p})}$ is the energy of the 1s2s,p states of the He atom, and $E_{i,\,\mathrm{He}}$ is the He ionization potential. Clearly, the 1s2s\,$^1S$ state is the dominant state producing ICD electrons. The 1s2s\,$^3S$ state and He$_2^*$ excimer states (broad feature around $E_{e,\, \mathrm{ICD}} =11$\,eV) also contribute but to a lesser extent. Although this electron spectrum was measured at a different excitation energy than those in the main text, ICD electrons appear to originate mostly from the same He$^*$ states. This is due to fast electronic relaxation, as previously observed in experiments using high-harmonic laser radiation~\cite{Ziemkiewicz:2015}, FEL~\cite{Mudrich2020,Ovcharenko2020} and synchrotron radiation~\cite{Buchta2013,BenLtaief2019}.

\begin{figure}
		\includegraphics[width=\linewidth]{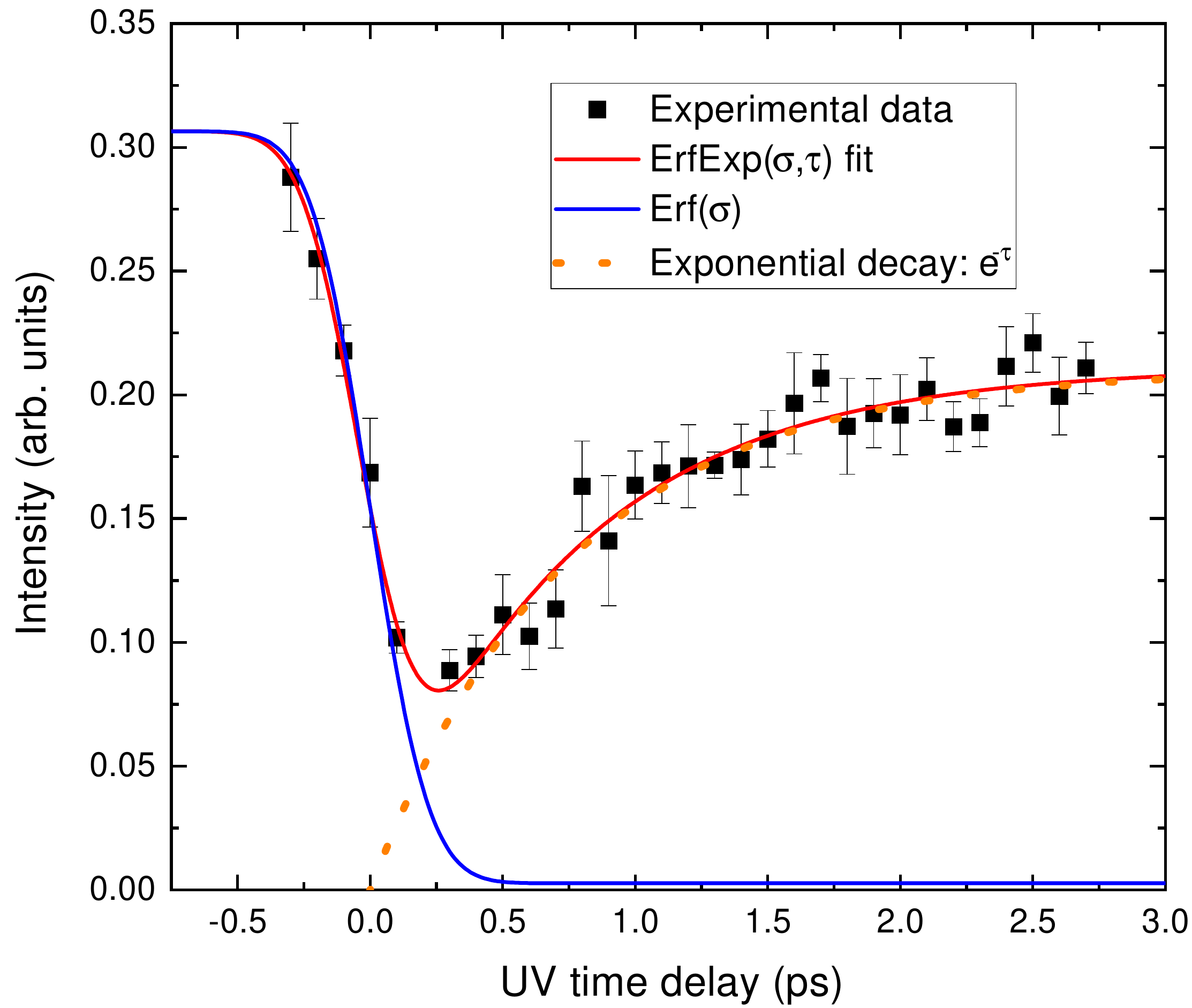}\caption{Fitting of experimental time-resolved ICD data (black squares) with the function from equation~\ref{eq:erfexp} (red curve). The components of the fit are illustrated separately: Error function (blue curve) and exponential decay (orange dotted line).} \label{fig:ICD:ErfExpFit}
\end{figure}

\section*{Appendix 2: Fitting of ICD electron yields}
The time-dependent ICD electron intensities are fitted with a convolution of the gaussian instrument response function obtained by resonant two-photon ionization of He and an exponential decay leading to the following function:
\begin{equation}
\label{eq:erfexp}
 \begin{split}
        I(t) = I_0 & -A \erfc\left[(\sigma^2-\tau(t-t_0))/(\sqrt{2}\sigma\tau)\right]\times \\ & \times\exp(-(t-t_0)/\tau)- B\erfc\left[ (t-t_0)/(\sqrt{2}\sigma)\right]
    \end{split}
\end{equation}

This model is the simplest analytic function that reproduces the experimental measurements. The exponential function reproduces the rise of the electron counts for long delay times. Thus, the exponential decay constant, $\tau$, represents the effective ICD time. The parameter $\sigma$ represents the cross-correlation width of the two overlapping laser pulses and was fixed to the value measured by resonant two-photon ionization of He gas. The time-zero value $t_0$ was constrained to $0\pm 15$~fs in order to account for possible drifts in the FEL timing. The free parameters $I_0$, $A$ and $B$ control the total ICD intensity for $t\rightarrow -\infty$, $t\rightarrow \infty$ and the maximum depletion $I_{min}$. 

Fig.~\ref{fig:ICD:ErfExpFit} displays a fit of a typical experimental measurement. In addition to the full fit curve (red), we show the separate contributions from the error function (blue line) and exponential decay (orange dots).

\begin{figure}
		\includegraphics[width=\linewidth]{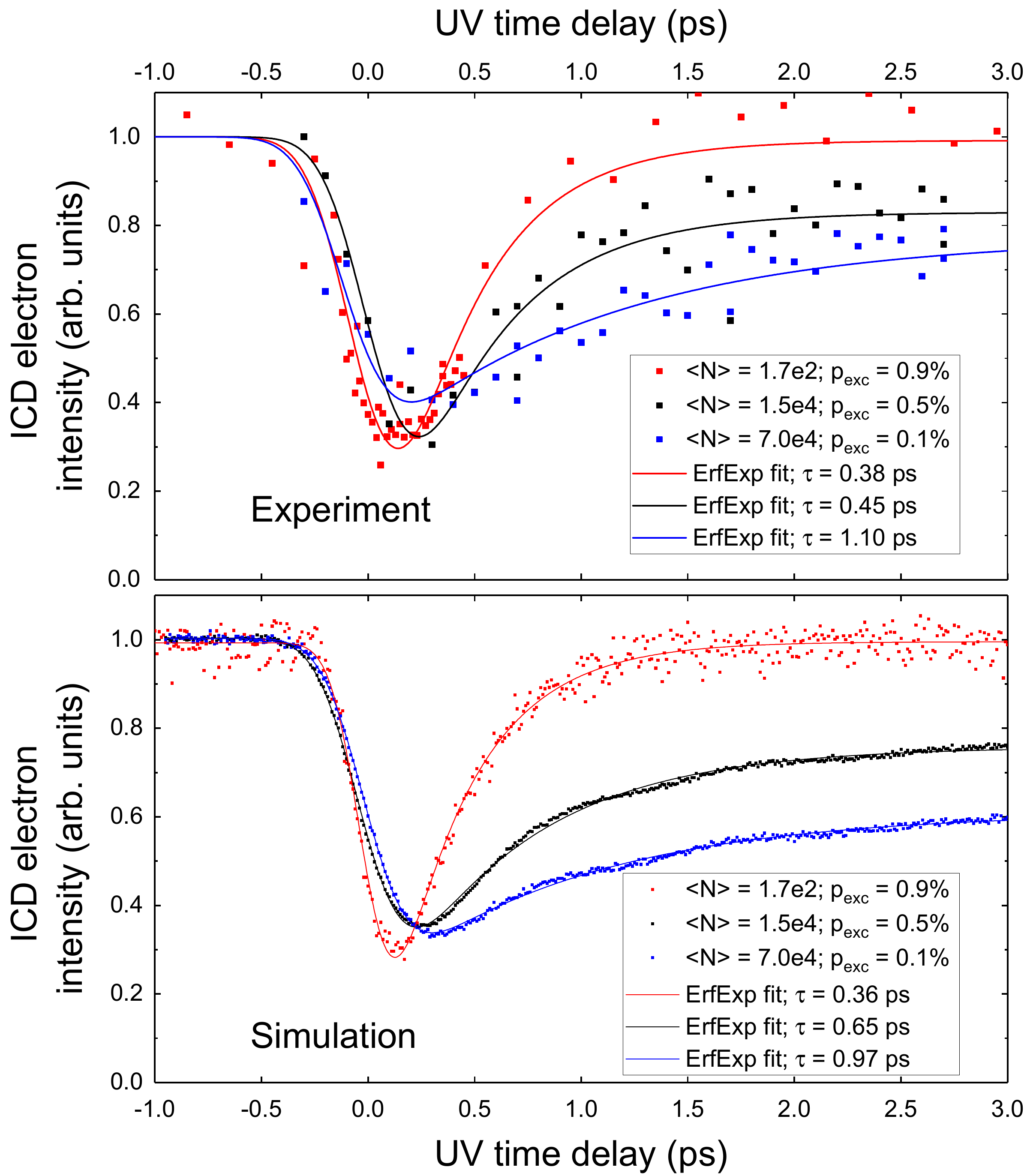}\caption{Upper: Experimental ICD electron intensities as a function of XUV-UV pump-probe delay with corresponding exponential fits for three different droplet sizes and excitation densities. Lower: Simulated ICD electron intensities and fits for conditions similar to those of the upper panel.} \label{fig:ICD:AddCurve}
	\end{figure}
	
\section*{Appendix 3: Additional experimental data}	
To give a better overview of the experimental results and systematics, we show in the upper panel of Fig.~\ref{fig:ICD:AddCurve} additional pump-probe ICD electron yields measured under different experimental conditions. The red symbols correspond to small droplets with high excitation density. The resulting ICD curve is characterized by a fast time variation as the mean interatomic distance between excited atoms is small,  $d<10$~\AA, and thus ICD is fast. The black curve is for an intermediate excitation density and intermediate droplet sizes. The blue curve is for large droplets combined with a low excitation density. Replenishment of the ICD electron signal after depletion is slower as ICD mostly occurs for pairs of He$^*$ with larger initial separation. 
The lower panel of Fig.~\ref{fig:ICD:AddCurve} shows the results of the MC simulation for the same parameters as in the experiment. The good agreement shows that our model captures the main aspects of the pump-probe ICD dynamics.

\begin{figure}
	\includegraphics[width=\linewidth]{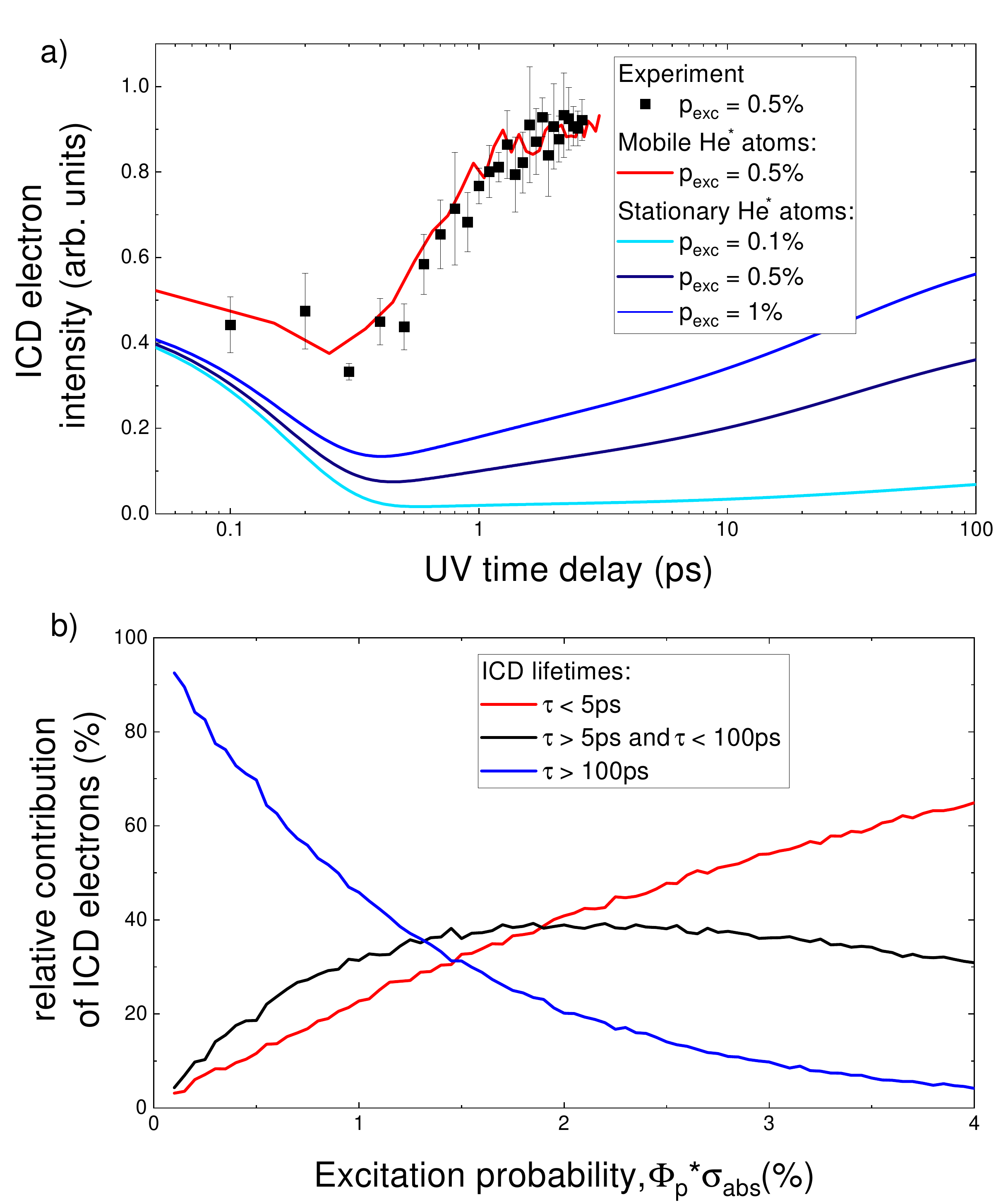}\caption{a) Simulated ICD electron intensities as a function of the UV time delay for fixed He$^*$ positions with different excitation densities (blue curves). For comparison, the experimental data is shown as black squares and the corresponding MC simulation assuming mobile He$^*$ atoms is shown as a red line. b) Relative contribution of ICD electrons broken into three different ICD lifetime intervals as a function of excitation probability for fixed He$^*$ positions.} \label{fig:ICDsim_verlaufe_tau}
\end{figure}

\section*{Appendix 4: The effect of atomic mobility on ICD timescales}
Besides providing a deeper understanding of our experimental findings, MC simulations additionally allow us to ask more fundamental questions about the process, which cannot be directly addressed through experiment. For instance, how important is the mobility of the He$^*$ atoms in the ICD process? To benchmark our simulations against the model system where the ICD rate is entirely given by the initial distances between He$^*$ we have carried out simulations where the He$^*$ positions are held fixed. Fig.~\ref{fig:ICDsim_verlaufe_tau} a) shows the simulated ICD electron intensity for stationary He$^*$ atoms as a function of the UV time delay for three different excitation probabilities (blue lines). For comparison, the experimental data is shown as black squares and the corresponding MC simulation assuming mobile He$^*$ atoms is shown as a red line. As can be clearly seen, the simulated dynamics for fixed He$^*$ positions proceed on much longer timescales compared to the experimental data, thus showing the critical importance of atomic mobility in the ICD process. To further illustrate this point, Fig.\,\ref{fig:ICDsim_verlaufe_tau} b) shows the relative contribution of ICD electrons broken into three different ICD lifetime intervals as a function of excitation probability for fixed He$^*$ positions. For low excitation probability ($\lesssim 1\,\%$,), the ICD lifetime would primarily be $\tau>100$\,ps, which is dramatically longer than what was shown in Fig. 2 for similar experimental conditions. Only for high excitation probability ($\gtrsim 2\,\%$) do shorter ICD lifetimes ($\tau<5$\,ps) significantly contribute.  We note that at $2.5\,\%$ excitation probability the transition from ICD to CAI occurs, in agreement with previous findings~\cite{Ovcharenko2020}.

\bibliography{wata1,papers2}

\begin{thebibliography}{49}%
\makeatletter
\providecommand \@ifxundefined [1]{%
 \@ifx{#1\undefined}
}%
\providecommand \@ifnum [1]{%
 \ifnum #1\expandafter \@firstoftwo
 \else \expandafter \@secondoftwo
 \fi
}%
\providecommand \@ifx [1]{%
 \ifx #1\expandafter \@firstoftwo
 \else \expandafter \@secondoftwo
 \fi
}%
\providecommand \natexlab [1]{#1}%
\providecommand \enquote  [1]{``#1''}%
\providecommand \bibnamefont  [1]{#1}%
\providecommand \bibfnamefont [1]{#1}%
\providecommand \citenamefont [1]{#1}%
\providecommand \href@noop [0]{\@secondoftwo}%
\providecommand \href [0]{\begingroup \@sanitize@url \@href}%
\providecommand \@href[1]{\@@startlink{#1}\@@href}%
\providecommand \@@href[1]{\endgroup#1\@@endlink}%
\providecommand \@sanitize@url [0]{\catcode `\\12\catcode `\$12\catcode
  `\&12\catcode `\#12\catcode `\^12\catcode `\_12\catcode `\%12\relax}%
\providecommand \@@startlink[1]{}%
\providecommand \@@endlink[0]{}%
\providecommand \url  [0]{\begingroup\@sanitize@url \@url }%
\providecommand \@url [1]{\endgroup\@href {#1}{\urlprefix }}%
\providecommand \urlprefix  [0]{URL }%
\providecommand \Eprint [0]{\href }%
\providecommand \doibase [0]{http://dx.doi.org/}%
\providecommand \selectlanguage [0]{\@gobble}%
\providecommand \bibinfo  [0]{\@secondoftwo}%
\providecommand \bibfield  [0]{\@secondoftwo}%
\providecommand \translation [1]{[#1]}%
\providecommand \BibitemOpen [0]{}%
\providecommand \bibitemStop [0]{}%
\providecommand \bibitemNoStop [0]{.\EOS\space}%
\providecommand \EOS [0]{\spacefactor3000\relax}%
\providecommand \BibitemShut  [1]{\csname bibitem#1\endcsname}%
\let\auto@bib@innerbib\@empty
\bibitem [{\citenamefont {Wabnitz}\ \emph {et~al.}(2002)\citenamefont
  {Wabnitz}, \citenamefont {Bittner}, \citenamefont {De~Castro}, \citenamefont
  {D{\"o}hrmann}, \citenamefont {G{\"u}rtler}, \citenamefont {Laarmann},
  \citenamefont {Laasch}, \citenamefont {Schulz}, \citenamefont {Swiderski},
  \citenamefont {von Haeften} \emph {et~al.}}]{Wabnitz2002}%
  \BibitemOpen
  \bibfield  {author} {\bibinfo {author} {\bibfnamefont {H.}~\bibnamefont
  {Wabnitz}}, \bibinfo {author} {\bibfnamefont {L.}~\bibnamefont {Bittner}},
  \bibinfo {author} {\bibfnamefont {A.}~\bibnamefont {De~Castro}}, \bibinfo
  {author} {\bibfnamefont {R.}~\bibnamefont {D{\"o}hrmann}}, \bibinfo {author}
  {\bibfnamefont {P.}~\bibnamefont {G{\"u}rtler}}, \bibinfo {author}
  {\bibfnamefont {T.}~\bibnamefont {Laarmann}}, \bibinfo {author}
  {\bibfnamefont {W.}~\bibnamefont {Laasch}}, \bibinfo {author} {\bibfnamefont
  {J.}~\bibnamefont {Schulz}}, \bibinfo {author} {\bibfnamefont
  {A.}~\bibnamefont {Swiderski}}, \bibinfo {author} {\bibfnamefont
  {K.}~\bibnamefont {von Haeften}},  \emph {et~al.},\ }\href@noop {} {\bibfield
   {journal} {\bibinfo  {journal} {Nature}\ }\textbf {\bibinfo {volume}
  {420}},\ \bibinfo {pages} {482} (\bibinfo {year} {2002})}\BibitemShut
  {NoStop}%
\bibitem [{\citenamefont {Bostedt}\ \emph {et~al.}(2008)\citenamefont
  {Bostedt}, \citenamefont {Thomas}, \citenamefont {Hoener}, \citenamefont
  {Eremina}, \citenamefont {Fennel}, \citenamefont {Meiwes-Broer},
  \citenamefont {Wabnitz}, \citenamefont {Kuhlmann}, \citenamefont
  {Pl{\"o}njes}, \citenamefont {Tiedtke} \emph {et~al.}}]{Bostedt2008}%
  \BibitemOpen
  \bibfield  {author} {\bibinfo {author} {\bibfnamefont {C.}~\bibnamefont
  {Bostedt}}, \bibinfo {author} {\bibfnamefont {H.}~\bibnamefont {Thomas}},
  \bibinfo {author} {\bibfnamefont {M.}~\bibnamefont {Hoener}}, \bibinfo
  {author} {\bibfnamefont {E.}~\bibnamefont {Eremina}}, \bibinfo {author}
  {\bibfnamefont {T.}~\bibnamefont {Fennel}}, \bibinfo {author} {\bibfnamefont
  {K.-H.}\ \bibnamefont {Meiwes-Broer}}, \bibinfo {author} {\bibfnamefont
  {H.}~\bibnamefont {Wabnitz}}, \bibinfo {author} {\bibfnamefont
  {M.}~\bibnamefont {Kuhlmann}}, \bibinfo {author} {\bibfnamefont
  {E.}~\bibnamefont {Pl{\"o}njes}}, \bibinfo {author} {\bibfnamefont
  {K.}~\bibnamefont {Tiedtke}},  \emph {et~al.},\ }\href@noop {} {\bibfield
  {journal} {\bibinfo  {journal} {Phys. Rev. Lett.}\ }\textbf {\bibinfo
  {volume} {100}},\ \bibinfo {pages} {133401} (\bibinfo {year}
  {2008})}\BibitemShut {NoStop}%
\bibitem [{\citenamefont {Bostedt}\ \emph {et~al.}(2010)\citenamefont
  {Bostedt}, \citenamefont {Thomas}, \citenamefont {Hoener}, \citenamefont
  {M{\"o}ller}, \citenamefont {Saalmann}, \citenamefont {Georgescu},
  \citenamefont {Gnodtke},\ and\ \citenamefont {Rost}}]{Bostedt2010}%
  \BibitemOpen
  \bibfield  {author} {\bibinfo {author} {\bibfnamefont {C.}~\bibnamefont
  {Bostedt}}, \bibinfo {author} {\bibfnamefont {H.}~\bibnamefont {Thomas}},
  \bibinfo {author} {\bibfnamefont {M.}~\bibnamefont {Hoener}}, \bibinfo
  {author} {\bibfnamefont {T.}~\bibnamefont {M{\"o}ller}}, \bibinfo {author}
  {\bibfnamefont {U.}~\bibnamefont {Saalmann}}, \bibinfo {author}
  {\bibfnamefont {I.}~\bibnamefont {Georgescu}}, \bibinfo {author}
  {\bibfnamefont {C.}~\bibnamefont {Gnodtke}}, \ and\ \bibinfo {author}
  {\bibfnamefont {J.-M.}\ \bibnamefont {Rost}},\ }\href@noop {} {\bibfield
  {journal} {\bibinfo  {journal} {New J. Phys.}\ }\textbf {\bibinfo {volume}
  {12}},\ \bibinfo {pages} {083004} (\bibinfo {year} {2010})}\BibitemShut
  {NoStop}%
\bibitem [{\citenamefont {Cederbaum}\ \emph {et~al.}(1997)\citenamefont
  {Cederbaum}, \citenamefont {Zobeley},\ and\ \citenamefont
  {Tarantelli}}]{Cederbaum1997}%
  \BibitemOpen
  \bibfield  {author} {\bibinfo {author} {\bibfnamefont {L.~S.}\ \bibnamefont
  {Cederbaum}}, \bibinfo {author} {\bibfnamefont {J.}~\bibnamefont {Zobeley}},
  \ and\ \bibinfo {author} {\bibfnamefont {F.}~\bibnamefont {Tarantelli}},\
  }\href {\doibase 10.1103/PhysRevLett.79.4778} {\bibfield  {journal} {\bibinfo
   {journal} {Phys. Rev. Lett.}\ }\textbf {\bibinfo {volume} {79}},\ \bibinfo
  {pages} {4778} (\bibinfo {year} {1997})}\BibitemShut {NoStop}%
\bibitem [{\citenamefont {Hergenhahn}(2011)}]{Hergenhahn2011}%
  \BibitemOpen
  \bibfield  {author} {\bibinfo {author} {\bibfnamefont {U.}~\bibnamefont
  {Hergenhahn}},\ }\href@noop {} {\bibfield  {journal} {\bibinfo  {journal} {J.
  Electron. Spectrosc. Relat. Phenom.}\ }\textbf {\bibinfo {volume} {184}},\
  \bibinfo {pages} {78} (\bibinfo {year} {2011})}\BibitemShut {NoStop}%
\bibitem [{\citenamefont {Jahnke}(2015)}]{Jahnke2015}%
  \BibitemOpen
  \bibfield  {author} {\bibinfo {author} {\bibfnamefont {T.}~\bibnamefont
  {Jahnke}},\ }\href@noop {} {\bibfield  {journal} {\bibinfo  {journal} {J.
  Phys. B: At., Mol. Opt. Phys.}\ }\textbf {\bibinfo {volume} {48}},\ \bibinfo
  {pages} {082001} (\bibinfo {year} {2015})}\BibitemShut {NoStop}%
\bibitem [{\citenamefont {Gokhberg}\ \emph {et~al.}(2014)\citenamefont
  {Gokhberg}, \citenamefont {Koloren{\v{c}}}, \citenamefont {Kuleff},\ and\
  \citenamefont {Cederbaum}}]{Gokhberg2014}%
  \BibitemOpen
  \bibfield  {author} {\bibinfo {author} {\bibfnamefont {K.}~\bibnamefont
  {Gokhberg}}, \bibinfo {author} {\bibfnamefont {P.}~\bibnamefont
  {Koloren{\v{c}}}}, \bibinfo {author} {\bibfnamefont {A.~I.}\ \bibnamefont
  {Kuleff}}, \ and\ \bibinfo {author} {\bibfnamefont {L.~S.}\ \bibnamefont
  {Cederbaum}},\ }\href@noop {} {\bibfield  {journal} {\bibinfo  {journal}
  {Nature}\ }\textbf {\bibinfo {volume} {505}},\ \bibinfo {pages} {661}
  (\bibinfo {year} {2014})}\BibitemShut {NoStop}%
\bibitem [{\citenamefont {Trinter}\ \emph {et~al.}(2014)\citenamefont
  {Trinter}, \citenamefont {Sch{\"o}ffler}, \citenamefont {Kim}, \citenamefont
  {Sturm}, \citenamefont {Cole}, \citenamefont {Neumann}, \citenamefont
  {Vredenborg}, \citenamefont {Williams}, \citenamefont {Bocharova},
  \citenamefont {Guillemin} \emph {et~al.}}]{Trinter2014}%
  \BibitemOpen
  \bibfield  {author} {\bibinfo {author} {\bibfnamefont {F.}~\bibnamefont
  {Trinter}}, \bibinfo {author} {\bibfnamefont {M.}~\bibnamefont
  {Sch{\"o}ffler}}, \bibinfo {author} {\bibfnamefont {H.-K.}\ \bibnamefont
  {Kim}}, \bibinfo {author} {\bibfnamefont {F.}~\bibnamefont {Sturm}}, \bibinfo
  {author} {\bibfnamefont {K.}~\bibnamefont {Cole}}, \bibinfo {author}
  {\bibfnamefont {N.}~\bibnamefont {Neumann}}, \bibinfo {author} {\bibfnamefont
  {A.}~\bibnamefont {Vredenborg}}, \bibinfo {author} {\bibfnamefont
  {J.}~\bibnamefont {Williams}}, \bibinfo {author} {\bibfnamefont
  {I.}~\bibnamefont {Bocharova}}, \bibinfo {author} {\bibfnamefont
  {R.}~\bibnamefont {Guillemin}},  \emph {et~al.},\ }\href@noop {} {\bibfield
  {journal} {\bibinfo  {journal} {Nature}\ }\textbf {\bibinfo {volume} {505}},\
  \bibinfo {pages} {664} (\bibinfo {year} {2014})}\BibitemShut {NoStop}%
\bibitem [{\citenamefont {Stumpf}\ \emph {et~al.}(2016)\citenamefont {Stumpf},
  \citenamefont {Gokhberg},\ and\ \citenamefont {Cederbaum}}]{Stumpf2016}%
  \BibitemOpen
  \bibfield  {author} {\bibinfo {author} {\bibfnamefont {V.}~\bibnamefont
  {Stumpf}}, \bibinfo {author} {\bibfnamefont {K.}~\bibnamefont {Gokhberg}}, \
  and\ \bibinfo {author} {\bibfnamefont {L.~S.}\ \bibnamefont {Cederbaum}},\
  }\href@noop {} {\bibfield  {journal} {\bibinfo  {journal} {Nat. Chem.}\
  }\textbf {\bibinfo {volume} {8}},\ \bibinfo {pages} {237} (\bibinfo {year}
  {2016})}\BibitemShut {NoStop}%
\bibitem [{\citenamefont {Ren}\ \emph {et~al.}(2018)\citenamefont {Ren},
  \citenamefont {Wang}, \citenamefont {Skitnevskaya}, \citenamefont {Trofimov},
  \citenamefont {Gokhberg},\ and\ \citenamefont {Dorn}}]{Ren2018}%
  \BibitemOpen
  \bibfield  {author} {\bibinfo {author} {\bibfnamefont {X.}~\bibnamefont
  {Ren}}, \bibinfo {author} {\bibfnamefont {E.}~\bibnamefont {Wang}}, \bibinfo
  {author} {\bibfnamefont {A.~D.}\ \bibnamefont {Skitnevskaya}}, \bibinfo
  {author} {\bibfnamefont {A.~B.}\ \bibnamefont {Trofimov}}, \bibinfo {author}
  {\bibfnamefont {K.}~\bibnamefont {Gokhberg}}, \ and\ \bibinfo {author}
  {\bibfnamefont {A.}~\bibnamefont {Dorn}},\ }\href@noop {} {\bibfield
  {journal} {\bibinfo  {journal} {Nat. Phys.}\ }\textbf {\bibinfo {volume}
  {14}},\ \bibinfo {pages} {1062} (\bibinfo {year} {2018})}\BibitemShut
  {NoStop}%
\bibitem [{\citenamefont {Allaria}\ \emph
  {et~al.}(2012{\natexlab{a}})\citenamefont {Allaria}, \citenamefont {Appio},
  \citenamefont {Badano}, \citenamefont {Barletta}, \citenamefont {Bassanese},
  \citenamefont {Biedron}, \citenamefont {Borga}, \citenamefont {Busetto},
  \citenamefont {Castronovo},\ and\ \citenamefont {Cinquegrana}}]{Allaria2012}%
  \BibitemOpen
  \bibfield  {author} {\bibinfo {author} {\bibfnamefont {E.}~\bibnamefont
  {Allaria}}, \bibinfo {author} {\bibfnamefont {R.}~\bibnamefont {Appio}},
  \bibinfo {author} {\bibfnamefont {L.}~\bibnamefont {Badano}}, \bibinfo
  {author} {\bibfnamefont {W.}~\bibnamefont {Barletta}}, \bibinfo {author}
  {\bibfnamefont {S.}~\bibnamefont {Bassanese}}, \bibinfo {author}
  {\bibfnamefont {S.}~\bibnamefont {Biedron}}, \bibinfo {author} {\bibfnamefont
  {A.}~\bibnamefont {Borga}}, \bibinfo {author} {\bibfnamefont
  {E.}~\bibnamefont {Busetto}}, \bibinfo {author} {\bibfnamefont
  {D.}~\bibnamefont {Castronovo}}, \ and\ \bibinfo {author} {\bibfnamefont
  {P.~e.~a.}\ \bibnamefont {Cinquegrana}},\ }\href@noop {} {\bibfield
  {journal} {\bibinfo  {journal} {Nat. Photonics}\ }\textbf {\bibinfo {volume}
  {6}},\ \bibinfo {pages} {699} (\bibinfo {year}
  {2012}{\natexlab{a}})}\BibitemShut {NoStop}%
\bibitem [{\citenamefont {Allaria}\ \emph
  {et~al.}(2012{\natexlab{b}})\citenamefont {Allaria}, \citenamefont
  {Battistoni}, \citenamefont {Bencivenga}, \citenamefont {Borghes},
  \citenamefont {Callegari}, \citenamefont {Capotondi}, \citenamefont
  {Castronovo}, \citenamefont {Cinquegrana}, \citenamefont {Cocco},
  \citenamefont {Coreno},\ and\ \citenamefont {et~al.}}]{Allaria2012a}%
  \BibitemOpen
  \bibfield  {author} {\bibinfo {author} {\bibfnamefont {E.}~\bibnamefont
  {Allaria}}, \bibinfo {author} {\bibfnamefont {A.}~\bibnamefont {Battistoni}},
  \bibinfo {author} {\bibfnamefont {F.}~\bibnamefont {Bencivenga}}, \bibinfo
  {author} {\bibfnamefont {R.}~\bibnamefont {Borghes}}, \bibinfo {author}
  {\bibfnamefont {C.}~\bibnamefont {Callegari}}, \bibinfo {author}
  {\bibfnamefont {F.}~\bibnamefont {Capotondi}}, \bibinfo {author}
  {\bibfnamefont {D.}~\bibnamefont {Castronovo}}, \bibinfo {author}
  {\bibfnamefont {P.}~\bibnamefont {Cinquegrana}}, \bibinfo {author}
  {\bibfnamefont {D.}~\bibnamefont {Cocco}}, \bibinfo {author} {\bibfnamefont
  {M.}~\bibnamefont {Coreno}}, \ and\ \bibinfo {author} {\bibnamefont
  {et~al.}},\ }\href@noop {} {\bibfield  {journal} {\bibinfo  {journal} {New J.
  Phys.}\ }\textbf {\bibinfo {volume} {14}},\ \bibinfo {pages} {113009}
  (\bibinfo {year} {2012}{\natexlab{b}})}\BibitemShut {NoStop}%
\bibitem [{\citenamefont {Kuleff}\ \emph {et~al.}(2010)\citenamefont {Kuleff},
  \citenamefont {Gokhberg}, \citenamefont {Kopelke},\ and\ \citenamefont
  {Cederbaum}}]{Kuleff2010}%
  \BibitemOpen
  \bibfield  {author} {\bibinfo {author} {\bibfnamefont {A.~I.}\ \bibnamefont
  {Kuleff}}, \bibinfo {author} {\bibfnamefont {K.}~\bibnamefont {Gokhberg}},
  \bibinfo {author} {\bibfnamefont {S.}~\bibnamefont {Kopelke}}, \ and\
  \bibinfo {author} {\bibfnamefont {L.~S.}\ \bibnamefont {Cederbaum}},\
  }\href@noop {} {\bibfield  {journal} {\bibinfo  {journal} {Phys. Rev. Lett.}\
  }\textbf {\bibinfo {volume} {105}},\ \bibinfo {pages} {043004} (\bibinfo
  {year} {2010})}\BibitemShut {NoStop}%
\bibitem [{\citenamefont {LaForge}\ \emph {et~al.}(2014)\citenamefont
  {LaForge}, \citenamefont {Drabbels}, \citenamefont {Brauer}, \citenamefont
  {Coreno}, \citenamefont {Devetta}, \citenamefont {Di~Fraia}, \citenamefont
  {Finetti}, \citenamefont {Grazioli}, \citenamefont {Katzy}, \citenamefont
  {Lyamayev} \emph {et~al.}}]{LaForge2014}%
  \BibitemOpen
  \bibfield  {author} {\bibinfo {author} {\bibfnamefont {A.}~\bibnamefont
  {LaForge}}, \bibinfo {author} {\bibfnamefont {M.}~\bibnamefont {Drabbels}},
  \bibinfo {author} {\bibfnamefont {N.~B.}\ \bibnamefont {Brauer}}, \bibinfo
  {author} {\bibfnamefont {M.}~\bibnamefont {Coreno}}, \bibinfo {author}
  {\bibfnamefont {M.}~\bibnamefont {Devetta}}, \bibinfo {author} {\bibfnamefont
  {M.}~\bibnamefont {Di~Fraia}}, \bibinfo {author} {\bibfnamefont
  {P.}~\bibnamefont {Finetti}}, \bibinfo {author} {\bibfnamefont
  {C.}~\bibnamefont {Grazioli}}, \bibinfo {author} {\bibfnamefont
  {R.}~\bibnamefont {Katzy}}, \bibinfo {author} {\bibfnamefont
  {V.}~\bibnamefont {Lyamayev}},  \emph {et~al.},\ }\href@noop {} {\bibfield
  {journal} {\bibinfo  {journal} {Sci. Rep.}\ }\textbf {\bibinfo {volume}
  {4}},\ \bibinfo {pages} {3621} (\bibinfo {year} {2014})}\BibitemShut
  {NoStop}%
\bibitem [{\citenamefont {Ovcharenko}\ \emph {et~al.}(2014)\citenamefont
  {Ovcharenko}, \citenamefont {Lyamayev}, \citenamefont {Katzy}, \citenamefont
  {Devetta}, \citenamefont {LaForge}, \citenamefont {O'Keeffe}, \citenamefont
  {Plekan}, \citenamefont {Finetti}, \citenamefont {Di~Fraia}, \citenamefont
  {Mudrich} \emph {et~al.}}]{Ovcharenko2014}%
  \BibitemOpen
  \bibfield  {author} {\bibinfo {author} {\bibfnamefont {Y.}~\bibnamefont
  {Ovcharenko}}, \bibinfo {author} {\bibfnamefont {V.}~\bibnamefont
  {Lyamayev}}, \bibinfo {author} {\bibfnamefont {R.}~\bibnamefont {Katzy}},
  \bibinfo {author} {\bibfnamefont {M.}~\bibnamefont {Devetta}}, \bibinfo
  {author} {\bibfnamefont {A.}~\bibnamefont {LaForge}}, \bibinfo {author}
  {\bibfnamefont {P.}~\bibnamefont {O'Keeffe}}, \bibinfo {author}
  {\bibfnamefont {O.}~\bibnamefont {Plekan}}, \bibinfo {author} {\bibfnamefont
  {P.}~\bibnamefont {Finetti}}, \bibinfo {author} {\bibfnamefont
  {M.}~\bibnamefont {Di~Fraia}}, \bibinfo {author} {\bibfnamefont
  {M.}~\bibnamefont {Mudrich}},  \emph {et~al.},\ }\href@noop {} {\bibfield
  {journal} {\bibinfo  {journal} {Phys. Rev. Lett.}\ }\textbf {\bibinfo
  {volume} {112}},\ \bibinfo {pages} {073401} (\bibinfo {year}
  {2014})}\BibitemShut {NoStop}%
\bibitem [{\citenamefont {Iablonskyi}\ \emph {et~al.}(2016)\citenamefont
  {Iablonskyi}, \citenamefont {Nagaya}, \citenamefont {Fukuzawa}, \citenamefont
  {Motomura}, \citenamefont {Kumagai}, \citenamefont {Mondal}, \citenamefont
  {Tachibana}, \citenamefont {Takanashi}, \citenamefont {Nishiyama},
  \citenamefont {Matsunami} \emph {et~al.}}]{Iablonskyi2016}%
  \BibitemOpen
  \bibfield  {author} {\bibinfo {author} {\bibfnamefont {D.}~\bibnamefont
  {Iablonskyi}}, \bibinfo {author} {\bibfnamefont {K.}~\bibnamefont {Nagaya}},
  \bibinfo {author} {\bibfnamefont {H.}~\bibnamefont {Fukuzawa}}, \bibinfo
  {author} {\bibfnamefont {K.}~\bibnamefont {Motomura}}, \bibinfo {author}
  {\bibfnamefont {Y.}~\bibnamefont {Kumagai}}, \bibinfo {author} {\bibfnamefont
  {S.}~\bibnamefont {Mondal}}, \bibinfo {author} {\bibfnamefont
  {T.}~\bibnamefont {Tachibana}}, \bibinfo {author} {\bibfnamefont
  {T.}~\bibnamefont {Takanashi}}, \bibinfo {author} {\bibfnamefont
  {T.}~\bibnamefont {Nishiyama}}, \bibinfo {author} {\bibfnamefont
  {K.}~\bibnamefont {Matsunami}},  \emph {et~al.},\ }\href@noop {} {\bibfield
  {journal} {\bibinfo  {journal} {Phys. Rev. Lett.}\ }\textbf {\bibinfo
  {volume} {117}},\ \bibinfo {pages} {276806} (\bibinfo {year}
  {2016})}\BibitemShut {NoStop}%
\bibitem [{\citenamefont {Ovcharenko}\ \emph {et~al.}(2020)\citenamefont
  {Ovcharenko}, \citenamefont {LaForge}, \citenamefont {Langbehn},
  \citenamefont {Plekan}, \citenamefont {Cucini}, \citenamefont {Finetti},
  \citenamefont {O'Keeffe}, \citenamefont {Iablonskyi}, \citenamefont
  {Nishiyama}, \citenamefont {Ueda}, \citenamefont {Piseri}, \citenamefont
  {Fraia}, \citenamefont {Richter}, \citenamefont {Coreno}, \citenamefont
  {Callegari}, \citenamefont {Prince}, \citenamefont {Stienkemeier},
  \citenamefont {Möller},\ and\ \citenamefont {Mudrich}}]{Ovcharenko2020}%
  \BibitemOpen
  \bibfield  {author} {\bibinfo {author} {\bibfnamefont {Y.}~\bibnamefont
  {Ovcharenko}}, \bibinfo {author} {\bibfnamefont {A.~C.}\ \bibnamefont
  {LaForge}}, \bibinfo {author} {\bibfnamefont {B.}~\bibnamefont {Langbehn}},
  \bibinfo {author} {\bibfnamefont {O.}~\bibnamefont {Plekan}}, \bibinfo
  {author} {\bibfnamefont {R.}~\bibnamefont {Cucini}}, \bibinfo {author}
  {\bibfnamefont {P.}~\bibnamefont {Finetti}}, \bibinfo {author} {\bibfnamefont
  {P.}~\bibnamefont {O'Keeffe}}, \bibinfo {author} {\bibfnamefont
  {D.}~\bibnamefont {Iablonskyi}}, \bibinfo {author} {\bibfnamefont
  {T.}~\bibnamefont {Nishiyama}}, \bibinfo {author} {\bibfnamefont
  {K.}~\bibnamefont {Ueda}}, \bibinfo {author} {\bibfnamefont {P.}~\bibnamefont
  {Piseri}}, \bibinfo {author} {\bibfnamefont {M.~D.}\ \bibnamefont {Fraia}},
  \bibinfo {author} {\bibfnamefont {R.}~\bibnamefont {Richter}}, \bibinfo
  {author} {\bibfnamefont {M.}~\bibnamefont {Coreno}}, \bibinfo {author}
  {\bibfnamefont {C.}~\bibnamefont {Callegari}}, \bibinfo {author}
  {\bibfnamefont {K.~C.}\ \bibnamefont {Prince}}, \bibinfo {author}
  {\bibfnamefont {F.}~\bibnamefont {Stienkemeier}}, \bibinfo {author}
  {\bibfnamefont {T.}~\bibnamefont {Möller}}, \ and\ \bibinfo {author}
  {\bibfnamefont {M.}~\bibnamefont {Mudrich}},\ }\href {\doibase
  10.1088/1367-2630/ab9554} {\bibfield  {journal} {\bibinfo  {journal} {New J.
  Phys.}\ }\textbf {\bibinfo {volume} {22}},\ \bibinfo {pages} {083043}
  (\bibinfo {year} {2020})}\BibitemShut {NoStop}%
\bibitem [{\citenamefont {Trinter}\ \emph {et~al.}(2013)\citenamefont
  {Trinter}, \citenamefont {Williams}, \citenamefont {Weller}, \citenamefont
  {Waitz}, \citenamefont {Pitzer}, \citenamefont {Voigtsberger}, \citenamefont
  {Schober}, \citenamefont {Kastirke}, \citenamefont {M{\"u}ller},
  \citenamefont {Goihl} \emph {et~al.}}]{Trinter2013}%
  \BibitemOpen
  \bibfield  {author} {\bibinfo {author} {\bibfnamefont {F.}~\bibnamefont
  {Trinter}}, \bibinfo {author} {\bibfnamefont {J.}~\bibnamefont {Williams}},
  \bibinfo {author} {\bibfnamefont {M.}~\bibnamefont {Weller}}, \bibinfo
  {author} {\bibfnamefont {M.}~\bibnamefont {Waitz}}, \bibinfo {author}
  {\bibfnamefont {M.}~\bibnamefont {Pitzer}}, \bibinfo {author} {\bibfnamefont
  {J.}~\bibnamefont {Voigtsberger}}, \bibinfo {author} {\bibfnamefont
  {C.}~\bibnamefont {Schober}}, \bibinfo {author} {\bibfnamefont
  {G.}~\bibnamefont {Kastirke}}, \bibinfo {author} {\bibfnamefont
  {C.}~\bibnamefont {M{\"u}ller}}, \bibinfo {author} {\bibfnamefont
  {C.}~\bibnamefont {Goihl}},  \emph {et~al.},\ }\href@noop {} {\bibfield
  {journal} {\bibinfo  {journal} {Phys. Rev. Lett.}\ }\textbf {\bibinfo
  {volume} {111}},\ \bibinfo {pages} {233004} (\bibinfo {year}
  {2013})}\BibitemShut {NoStop}%
\bibitem [{\citenamefont {Buchta}\ \emph {et~al.}(2013)\citenamefont {Buchta},
  \citenamefont {Krishnan}, \citenamefont {Brauer}, \citenamefont {Drabbels},
  \citenamefont {O'Keeffe}, \citenamefont {Devetta}, \citenamefont {Di~Fraia},
  \citenamefont {Callegari}, \citenamefont {Richter}, \citenamefont {Coreno}
  \emph {et~al.}}]{Buchta2013}%
  \BibitemOpen
  \bibfield  {author} {\bibinfo {author} {\bibfnamefont {D.}~\bibnamefont
  {Buchta}}, \bibinfo {author} {\bibfnamefont {S.~R.}\ \bibnamefont
  {Krishnan}}, \bibinfo {author} {\bibfnamefont {N.~B.}\ \bibnamefont
  {Brauer}}, \bibinfo {author} {\bibfnamefont {M.}~\bibnamefont {Drabbels}},
  \bibinfo {author} {\bibfnamefont {P.}~\bibnamefont {O'Keeffe}}, \bibinfo
  {author} {\bibfnamefont {M.}~\bibnamefont {Devetta}}, \bibinfo {author}
  {\bibfnamefont {M.}~\bibnamefont {Di~Fraia}}, \bibinfo {author}
  {\bibfnamefont {C.}~\bibnamefont {Callegari}}, \bibinfo {author}
  {\bibfnamefont {R.}~\bibnamefont {Richter}}, \bibinfo {author} {\bibfnamefont
  {M.}~\bibnamefont {Coreno}},  \emph {et~al.},\ }\href@noop {} {\bibfield
  {journal} {\bibinfo  {journal} {J. Phys. Chem. A}\ }\textbf {\bibinfo
  {volume} {117}},\ \bibinfo {pages} {4394} (\bibinfo {year}
  {2013})}\BibitemShut {NoStop}%
\bibitem [{\citenamefont {LaForge}\ \emph {et~al.}(2019)\citenamefont
  {LaForge}, \citenamefont {Shcherbinin}, \citenamefont {Stienkemeier},
  \citenamefont {Richter}, \citenamefont {Moshammer}, \citenamefont {Pfeifer},\
  and\ \citenamefont {Mudrich}}]{LaForge2019}%
  \BibitemOpen
  \bibfield  {author} {\bibinfo {author} {\bibfnamefont {A.}~\bibnamefont
  {LaForge}}, \bibinfo {author} {\bibfnamefont {M.}~\bibnamefont
  {Shcherbinin}}, \bibinfo {author} {\bibfnamefont {F.}~\bibnamefont
  {Stienkemeier}}, \bibinfo {author} {\bibfnamefont {R.}~\bibnamefont
  {Richter}}, \bibinfo {author} {\bibfnamefont {R.}~\bibnamefont {Moshammer}},
  \bibinfo {author} {\bibfnamefont {T.}~\bibnamefont {Pfeifer}}, \ and\
  \bibinfo {author} {\bibfnamefont {M.}~\bibnamefont {Mudrich}},\ }\href@noop
  {} {\bibfield  {journal} {\bibinfo  {journal} {Nat. Phys.}\ }\textbf
  {\bibinfo {volume} {15}},\ \bibinfo {pages} {247} (\bibinfo {year}
  {2019})}\BibitemShut {NoStop}%
\bibitem [{\citenamefont {Havermeier}\ \emph {et~al.}(2010)\citenamefont
  {Havermeier}, \citenamefont {Jahnke}, \citenamefont {Kreidi}, \citenamefont
  {Wallauer}, \citenamefont {Voss}, \citenamefont {Sch{\"o}ffler},
  \citenamefont {Sch{\"o}ssler}, \citenamefont {Foucar}, \citenamefont
  {Neumann}, \citenamefont {Titze} \emph {et~al.}}]{Havermeier2010}%
  \BibitemOpen
  \bibfield  {author} {\bibinfo {author} {\bibfnamefont {T.}~\bibnamefont
  {Havermeier}}, \bibinfo {author} {\bibfnamefont {T.}~\bibnamefont {Jahnke}},
  \bibinfo {author} {\bibfnamefont {K.}~\bibnamefont {Kreidi}}, \bibinfo
  {author} {\bibfnamefont {R.}~\bibnamefont {Wallauer}}, \bibinfo {author}
  {\bibfnamefont {S.}~\bibnamefont {Voss}}, \bibinfo {author} {\bibfnamefont
  {M.}~\bibnamefont {Sch{\"o}ffler}}, \bibinfo {author} {\bibfnamefont
  {S.}~\bibnamefont {Sch{\"o}ssler}}, \bibinfo {author} {\bibfnamefont
  {L.}~\bibnamefont {Foucar}}, \bibinfo {author} {\bibfnamefont
  {N.}~\bibnamefont {Neumann}}, \bibinfo {author} {\bibfnamefont
  {J.}~\bibnamefont {Titze}},  \emph {et~al.},\ }\href@noop {} {\bibfield
  {journal} {\bibinfo  {journal} {Phys. Rev. Lett.}\ }\textbf {\bibinfo
  {volume} {104}},\ \bibinfo {pages} {133401} (\bibinfo {year}
  {2010})}\BibitemShut {NoStop}%
\bibitem [{\citenamefont {Sisourat}\ \emph {et~al.}(2010)\citenamefont
  {Sisourat}, \citenamefont {Kryzhevoi}, \citenamefont {Koloren{\v{c}}},
  \citenamefont {Scheit}, \citenamefont {Jahnke},\ and\ \citenamefont
  {Cederbaum}}]{Sisourat2010}%
  \BibitemOpen
  \bibfield  {author} {\bibinfo {author} {\bibfnamefont {N.}~\bibnamefont
  {Sisourat}}, \bibinfo {author} {\bibfnamefont {N.~V.}\ \bibnamefont
  {Kryzhevoi}}, \bibinfo {author} {\bibfnamefont {P.}~\bibnamefont
  {Koloren{\v{c}}}}, \bibinfo {author} {\bibfnamefont {S.}~\bibnamefont
  {Scheit}}, \bibinfo {author} {\bibfnamefont {T.}~\bibnamefont {Jahnke}}, \
  and\ \bibinfo {author} {\bibfnamefont {L.~S.}\ \bibnamefont {Cederbaum}},\
  }\href@noop {} {\bibfield  {journal} {\bibinfo  {journal} {Nat. Phys.}\
  }\textbf {\bibinfo {volume} {6}},\ \bibinfo {pages} {508} (\bibinfo {year}
  {2010})}\BibitemShut {NoStop}%
\bibitem [{\citenamefont {Stumpf}\ \emph {et~al.}(2014)\citenamefont {Stumpf},
  \citenamefont {Kryzhevoi}, \citenamefont {Gokhberg},\ and\ \citenamefont
  {Cederbaum}}]{Stumpf2014}%
  \BibitemOpen
  \bibfield  {author} {\bibinfo {author} {\bibfnamefont {V.}~\bibnamefont
  {Stumpf}}, \bibinfo {author} {\bibfnamefont {N.}~\bibnamefont {Kryzhevoi}},
  \bibinfo {author} {\bibfnamefont {K.}~\bibnamefont {Gokhberg}}, \ and\
  \bibinfo {author} {\bibfnamefont {L.}~\bibnamefont {Cederbaum}},\ }\href@noop
  {} {\bibfield  {journal} {\bibinfo  {journal} {Phys. Rev. Lett.}\ }\textbf
  {\bibinfo {volume} {112}},\ \bibinfo {pages} {193001} (\bibinfo {year}
  {2014})}\BibitemShut {NoStop}%
\bibitem [{\citenamefont {LaForge}\ \emph {et~al.}(2016)\citenamefont
  {LaForge}, \citenamefont {Stumpf}, \citenamefont {Gokhberg}, \citenamefont
  {von Vangerow}, \citenamefont {Stienkemeier}, \citenamefont {Kryzhevoi},
  \citenamefont {O'Keeffe}, \citenamefont {Ciavardini}, \citenamefont
  {Krishnan}, \citenamefont {Coreno} \emph {et~al.}}]{LaForge2016}%
  \BibitemOpen
  \bibfield  {author} {\bibinfo {author} {\bibfnamefont {A.}~\bibnamefont
  {LaForge}}, \bibinfo {author} {\bibfnamefont {V.}~\bibnamefont {Stumpf}},
  \bibinfo {author} {\bibfnamefont {K.}~\bibnamefont {Gokhberg}}, \bibinfo
  {author} {\bibfnamefont {J.}~\bibnamefont {von Vangerow}}, \bibinfo {author}
  {\bibfnamefont {F.}~\bibnamefont {Stienkemeier}}, \bibinfo {author}
  {\bibfnamefont {N.}~\bibnamefont {Kryzhevoi}}, \bibinfo {author}
  {\bibfnamefont {P.}~\bibnamefont {O'Keeffe}}, \bibinfo {author}
  {\bibfnamefont {A.}~\bibnamefont {Ciavardini}}, \bibinfo {author}
  {\bibfnamefont {S.}~\bibnamefont {Krishnan}}, \bibinfo {author}
  {\bibfnamefont {M.}~\bibnamefont {Coreno}},  \emph {et~al.},\ }\href@noop {}
  {\bibfield  {journal} {\bibinfo  {journal} {Phys. Rev. Lett.}\ }\textbf
  {\bibinfo {volume} {116}},\ \bibinfo {pages} {203001} (\bibinfo {year}
  {2016})}\BibitemShut {NoStop}%
\bibitem [{\citenamefont {Shcherbinin}\ \emph {et~al.}(2017)\citenamefont
  {Shcherbinin}, \citenamefont {LaForge}, \citenamefont {Sharma}, \citenamefont
  {Devetta}, \citenamefont {Richter}, \citenamefont {Moshammer}, \citenamefont
  {Pfeifer},\ and\ \citenamefont {Mudrich}}]{Shcherbinin2017}%
  \BibitemOpen
  \bibfield  {author} {\bibinfo {author} {\bibfnamefont {M.}~\bibnamefont
  {Shcherbinin}}, \bibinfo {author} {\bibfnamefont {A.}~\bibnamefont
  {LaForge}}, \bibinfo {author} {\bibfnamefont {V.}~\bibnamefont {Sharma}},
  \bibinfo {author} {\bibfnamefont {M.}~\bibnamefont {Devetta}}, \bibinfo
  {author} {\bibfnamefont {R.}~\bibnamefont {Richter}}, \bibinfo {author}
  {\bibfnamefont {R.}~\bibnamefont {Moshammer}}, \bibinfo {author}
  {\bibfnamefont {T.}~\bibnamefont {Pfeifer}}, \ and\ \bibinfo {author}
  {\bibfnamefont {M.}~\bibnamefont {Mudrich}},\ }\href@noop {} {\bibfield
  {journal} {\bibinfo  {journal} {Phys. Rev. A}\ }\textbf {\bibinfo {volume}
  {96}},\ \bibinfo {pages} {013407} (\bibinfo {year} {2017})}\BibitemShut
  {NoStop}%
\bibitem [{\citenamefont {Ben~Ltaief}\ \emph {et~al.}(2019)\citenamefont
  {Ben~Ltaief}, \citenamefont {Shcherbinin}, \citenamefont {Mandal},
  \citenamefont {Krishnan}, \citenamefont {LaForge}, \citenamefont {Richter},
  \citenamefont {Turchini}, \citenamefont {Zema}, \citenamefont {Pfeifer},
  \citenamefont {Fasshauer} \emph {et~al.}}]{BenLtaief2019}%
  \BibitemOpen
  \bibfield  {author} {\bibinfo {author} {\bibfnamefont {L.}~\bibnamefont
  {Ben~Ltaief}}, \bibinfo {author} {\bibfnamefont {M.}~\bibnamefont
  {Shcherbinin}}, \bibinfo {author} {\bibfnamefont {S.}~\bibnamefont {Mandal}},
  \bibinfo {author} {\bibfnamefont {S.}~\bibnamefont {Krishnan}}, \bibinfo
  {author} {\bibfnamefont {A.}~\bibnamefont {LaForge}}, \bibinfo {author}
  {\bibfnamefont {R.}~\bibnamefont {Richter}}, \bibinfo {author} {\bibfnamefont
  {S.}~\bibnamefont {Turchini}}, \bibinfo {author} {\bibfnamefont
  {N.}~\bibnamefont {Zema}}, \bibinfo {author} {\bibfnamefont {T.}~\bibnamefont
  {Pfeifer}}, \bibinfo {author} {\bibfnamefont {E.}~\bibnamefont {Fasshauer}},
  \emph {et~al.},\ }\href@noop {} {\bibfield  {journal} {\bibinfo  {journal}
  {J. Phys. Chem. Lett.}\ }\textbf {\bibinfo {volume} {10}},\ \bibinfo {pages}
  {6904} (\bibinfo {year} {2019})}\BibitemShut {NoStop}%
\bibitem [{\citenamefont {Ltaief}\ \emph {et~al.}(2020)\citenamefont {Ltaief},
  \citenamefont {Shcherbinin}, \citenamefont {Mandal}, \citenamefont
  {Krishnan}, \citenamefont {Richter}, \citenamefont {Pfeifer}, \citenamefont
  {Bauer}, \citenamefont {Ghosh}, \citenamefont {Mudrich}, \citenamefont
  {Gokhberg} \emph {et~al.}}]{BenLtaief2020}%
  \BibitemOpen
  \bibfield  {author} {\bibinfo {author} {\bibfnamefont {L.~B.}\ \bibnamefont
  {Ltaief}}, \bibinfo {author} {\bibfnamefont {M.}~\bibnamefont {Shcherbinin}},
  \bibinfo {author} {\bibfnamefont {S.}~\bibnamefont {Mandal}}, \bibinfo
  {author} {\bibfnamefont {S.~R.}\ \bibnamefont {Krishnan}}, \bibinfo {author}
  {\bibfnamefont {R.}~\bibnamefont {Richter}}, \bibinfo {author} {\bibfnamefont
  {T.}~\bibnamefont {Pfeifer}}, \bibinfo {author} {\bibfnamefont
  {M.}~\bibnamefont {Bauer}}, \bibinfo {author} {\bibfnamefont
  {A.}~\bibnamefont {Ghosh}}, \bibinfo {author} {\bibfnamefont
  {M.}~\bibnamefont {Mudrich}}, \bibinfo {author} {\bibfnamefont
  {K.}~\bibnamefont {Gokhberg}},  \emph {et~al.},\ }\href@noop {} {\bibfield
  {journal} {\bibinfo  {journal} {Phys. Chem. Chem. Phys.}\ }\textbf {\bibinfo
  {volume} {22}},\ \bibinfo {pages} {8557} (\bibinfo {year}
  {2020})}\BibitemShut {NoStop}%
\bibitem [{\citenamefont {Katzy}\ \emph {et~al.}(2015)\citenamefont {Katzy},
  \citenamefont {LaForge}, \citenamefont {Ovcharenko}, \citenamefont {Coreno},
  \citenamefont {Devetta}, \citenamefont {Di~Fraia}, \citenamefont {Drabbels},
  \citenamefont {Finetti}, \citenamefont {Lyamayev}, \citenamefont {Mazza}
  \emph {et~al.}}]{Katzy2015}%
  \BibitemOpen
  \bibfield  {author} {\bibinfo {author} {\bibfnamefont {R.}~\bibnamefont
  {Katzy}}, \bibinfo {author} {\bibfnamefont {A.}~\bibnamefont {LaForge}},
  \bibinfo {author} {\bibfnamefont {Y.}~\bibnamefont {Ovcharenko}}, \bibinfo
  {author} {\bibfnamefont {M.}~\bibnamefont {Coreno}}, \bibinfo {author}
  {\bibfnamefont {M.}~\bibnamefont {Devetta}}, \bibinfo {author} {\bibfnamefont
  {M.}~\bibnamefont {Di~Fraia}}, \bibinfo {author} {\bibfnamefont
  {M.}~\bibnamefont {Drabbels}}, \bibinfo {author} {\bibfnamefont
  {P.}~\bibnamefont {Finetti}}, \bibinfo {author} {\bibfnamefont
  {V.}~\bibnamefont {Lyamayev}}, \bibinfo {author} {\bibfnamefont
  {T.}~\bibnamefont {Mazza}},  \emph {et~al.},\ }\href@noop {} {\bibfield
  {journal} {\bibinfo  {journal} {J. Phys. B: At., Mol. Opt. Phys.}\ }\textbf
  {\bibinfo {volume} {48}},\ \bibinfo {pages} {244011} (\bibinfo {year}
  {2015})}\BibitemShut {NoStop}%
\bibitem [{\citenamefont {von Haeften}\ \emph {et~al.}(2002)\citenamefont {von
  Haeften}, \citenamefont {Laarmann}, \citenamefont {Wabnitz},\ and\
  \citenamefont {M{\"o}ller}}]{Haeften2002}%
  \BibitemOpen
  \bibfield  {author} {\bibinfo {author} {\bibfnamefont {K.}~\bibnamefont {von
  Haeften}}, \bibinfo {author} {\bibfnamefont {T.}~\bibnamefont {Laarmann}},
  \bibinfo {author} {\bibfnamefont {H.}~\bibnamefont {Wabnitz}}, \ and\
  \bibinfo {author} {\bibfnamefont {T.}~\bibnamefont {M{\"o}ller}},\
  }\href@noop {} {\bibfield  {journal} {\bibinfo  {journal} {Phys. Rev. Lett.}\
  }\textbf {\bibinfo {volume} {88}},\ \bibinfo {pages} {233401} (\bibinfo
  {year} {2002})}\BibitemShut {NoStop}%
\bibitem [{\citenamefont {Thaler}\ \emph {et~al.}(2018)\citenamefont {Thaler},
  \citenamefont {Ranftl}, \citenamefont {Heim}, \citenamefont {Cesnik},
  \citenamefont {Treiber}, \citenamefont {Meyer}, \citenamefont {Hauser},
  \citenamefont {Ernst},\ and\ \citenamefont {Koch}}]{Thaler2018}%
  \BibitemOpen
  \bibfield  {author} {\bibinfo {author} {\bibfnamefont {B.}~\bibnamefont
  {Thaler}}, \bibinfo {author} {\bibfnamefont {S.}~\bibnamefont {Ranftl}},
  \bibinfo {author} {\bibfnamefont {P.}~\bibnamefont {Heim}}, \bibinfo {author}
  {\bibfnamefont {S.}~\bibnamefont {Cesnik}}, \bibinfo {author} {\bibfnamefont
  {L.}~\bibnamefont {Treiber}}, \bibinfo {author} {\bibfnamefont
  {R.}~\bibnamefont {Meyer}}, \bibinfo {author} {\bibfnamefont {A.~W.}\
  \bibnamefont {Hauser}}, \bibinfo {author} {\bibfnamefont {W.~E.}\
  \bibnamefont {Ernst}}, \ and\ \bibinfo {author} {\bibfnamefont
  {M.}~\bibnamefont {Koch}},\ }\href@noop {} {\bibfield  {journal} {\bibinfo
  {journal} {Nat. Commun.}\ }\textbf {\bibinfo {volume} {9}},\ \bibinfo {pages}
  {1} (\bibinfo {year} {2018})}\BibitemShut {NoStop}%
\bibitem [{\citenamefont {Mudrich}\ \emph {et~al.}(2020)\citenamefont
  {Mudrich}, \citenamefont {LaForge}, \citenamefont {Ciavardini}, \citenamefont
  {OKeeffe}, \citenamefont {Callegari}, \citenamefont {Coreno}, \citenamefont
  {Demidovich}, \citenamefont {Devetta}, \citenamefont {Di~Fraia},
  \citenamefont {Drabbels} \emph {et~al.}}]{Mudrich2020}%
  \BibitemOpen
  \bibfield  {author} {\bibinfo {author} {\bibfnamefont {M.}~\bibnamefont
  {Mudrich}}, \bibinfo {author} {\bibfnamefont {A.}~\bibnamefont {LaForge}},
  \bibinfo {author} {\bibfnamefont {A.}~\bibnamefont {Ciavardini}}, \bibinfo
  {author} {\bibfnamefont {P.}~\bibnamefont {O'Keeffe}}, \bibinfo {author}
  {\bibfnamefont {C.}~\bibnamefont {Callegari}}, \bibinfo {author}
  {\bibfnamefont {M.}~\bibnamefont {Coreno}}, \bibinfo {author} {\bibfnamefont
  {A.}~\bibnamefont {Demidovich}}, \bibinfo {author} {\bibfnamefont
  {M.}~\bibnamefont {Devetta}}, \bibinfo {author} {\bibfnamefont
  {M.}~\bibnamefont {Di~Fraia}}, \bibinfo {author} {\bibfnamefont
  {M.}~\bibnamefont {Drabbels}},  \emph {et~al.},\ }\href@noop {} {\bibfield
  {journal} {\bibinfo  {journal} {Nat. Commun.}\ }\textbf {\bibinfo {volume}
  {11}},\ \bibinfo {pages} {112} (\bibinfo {year} {2020})}\BibitemShut
  {NoStop}%
\bibitem [{\citenamefont {Toennies}\ and\ \citenamefont
  {Vilesov}(2004)}]{Toennies2004}%
  \BibitemOpen
  \bibfield  {author} {\bibinfo {author} {\bibfnamefont {J.~P.}\ \bibnamefont
  {Toennies}}\ and\ \bibinfo {author} {\bibfnamefont {A.~F.}\ \bibnamefont
  {Vilesov}},\ }\href@noop {} {\bibfield  {journal} {\bibinfo  {journal}
  {Angew. Chem. Int. Ed.}\ }\textbf {\bibinfo {volume} {43}},\ \bibinfo {pages}
  {2622} (\bibinfo {year} {2004})}\BibitemShut {NoStop}%
\bibitem [{\citenamefont {Joppien}\ \emph {et~al.}(1993)\citenamefont
  {Joppien}, \citenamefont {Karnbach},\ and\ \citenamefont
  {M{\"o}ller}}]{Joppien1993}%
  \BibitemOpen
  \bibfield  {author} {\bibinfo {author} {\bibfnamefont {M.}~\bibnamefont
  {Joppien}}, \bibinfo {author} {\bibfnamefont {R.}~\bibnamefont {Karnbach}}, \
  and\ \bibinfo {author} {\bibfnamefont {T.}~\bibnamefont {M{\"o}ller}},\
  }\href@noop {} {\bibfield  {journal} {\bibinfo  {journal} {Phys. Rev. Lett.}\
  }\textbf {\bibinfo {volume} {71}},\ \bibinfo {pages} {2654} (\bibinfo {year}
  {1993})}\BibitemShut {NoStop}%
\bibitem [{\citenamefont {Ancilotto}\ \emph {et~al.}(2017)\citenamefont
  {Ancilotto} \emph {et~al.}}]{Ancilotto:2017}%
  \BibitemOpen
  \bibfield  {author} {\bibinfo {author} {\bibfnamefont {F.}~\bibnamefont
  {Ancilotto}} \emph {et~al.},\ }\href@noop {} {\bibfield  {journal} {\bibinfo
  {journal} {Int. Rev. Phys. Chem.}\ }\textbf {\bibinfo {volume} {36}},\
  \bibinfo {pages} {621} (\bibinfo {year} {2017})}\BibitemShut {NoStop}%
\bibitem [{\citenamefont {Lyamayev}\ \emph {et~al.}(2013)\citenamefont
  {Lyamayev}, \citenamefont {Ovcharenko}, \citenamefont {Katzy}, \citenamefont
  {Devetta}, \citenamefont {Bruder}, \citenamefont {LaForge}, \citenamefont
  {Mudrich}, \citenamefont {Person}, \citenamefont {Stienkemeier},
  \citenamefont {Krikunova},\ and\ \citenamefont {et~al.}}]{Lyamayev2013}%
  \BibitemOpen
  \bibfield  {author} {\bibinfo {author} {\bibfnamefont {V.}~\bibnamefont
  {Lyamayev}}, \bibinfo {author} {\bibfnamefont {Y.}~\bibnamefont
  {Ovcharenko}}, \bibinfo {author} {\bibfnamefont {R.}~\bibnamefont {Katzy}},
  \bibinfo {author} {\bibfnamefont {M.}~\bibnamefont {Devetta}}, \bibinfo
  {author} {\bibfnamefont {L.}~\bibnamefont {Bruder}}, \bibinfo {author}
  {\bibfnamefont {A.}~\bibnamefont {LaForge}}, \bibinfo {author} {\bibfnamefont
  {M.}~\bibnamefont {Mudrich}}, \bibinfo {author} {\bibfnamefont
  {U.}~\bibnamefont {Person}}, \bibinfo {author} {\bibfnamefont
  {F.}~\bibnamefont {Stienkemeier}}, \bibinfo {author} {\bibfnamefont
  {M.}~\bibnamefont {Krikunova}}, \ and\ \bibinfo {author} {\bibnamefont
  {et~al.}},\ }\href@noop {} {\bibfield  {journal} {\bibinfo  {journal} {J.
  Phys. B}\ }\textbf {\bibinfo {volume} {46}},\ \bibinfo {pages} {164007}
  (\bibinfo {year} {2013})}\BibitemShut {NoStop}%
\bibitem [{\citenamefont {Dick}(2014)}]{Dick2014}%
  \BibitemOpen
  \bibfield  {author} {\bibinfo {author} {\bibfnamefont {B.}~\bibnamefont
  {Dick}},\ }\href@noop {} {\bibfield  {journal} {\bibinfo  {journal} {Phys.
  Chem. Chem. Phys.}\ }\textbf {\bibinfo {volume} {16}},\ \bibinfo {pages}
  {570} (\bibinfo {year} {2014})}\BibitemShut {NoStop}%
\bibitem [{\citenamefont {Chang}\ and\ \citenamefont
  {Fang}(1995)}]{chang1995effect}%
  \BibitemOpen
  \bibfield  {author} {\bibinfo {author} {\bibfnamefont {T.}~\bibnamefont
  {Chang}}\ and\ \bibinfo {author} {\bibfnamefont {T.}~\bibnamefont {Fang}},\
  }\href@noop {} {\bibfield  {journal} {\bibinfo  {journal} {Phys. Rev. A}\
  }\textbf {\bibinfo {volume} {52}},\ \bibinfo {pages} {2638} (\bibinfo {year}
  {1995})}\BibitemShut {NoStop}%
\bibitem [{\citenamefont {Miteva}\ \emph {et~al.}(2017)\citenamefont {Miteva},
  \citenamefont {Kazandjian},\ and\ \citenamefont
  {Sisourat}}]{miteva2017computations}%
  \BibitemOpen
  \bibfield  {author} {\bibinfo {author} {\bibfnamefont {T.}~\bibnamefont
  {Miteva}}, \bibinfo {author} {\bibfnamefont {S.}~\bibnamefont {Kazandjian}},
  \ and\ \bibinfo {author} {\bibfnamefont {N.}~\bibnamefont {Sisourat}},\
  }\href@noop {} {\bibfield  {journal} {\bibinfo  {journal} {Chem. Phys.}\
  }\textbf {\bibinfo {volume} {482}},\ \bibinfo {pages} {208} (\bibinfo {year}
  {2017})}\BibitemShut {NoStop}%
\bibitem [{\citenamefont {Scharf}\ \emph {et~al.}(1986)\citenamefont {Scharf},
  \citenamefont {Jortner},\ and\ \citenamefont {Landman}}]{Scharf1986}%
  \BibitemOpen
  \bibfield  {author} {\bibinfo {author} {\bibfnamefont {D.}~\bibnamefont
  {Scharf}}, \bibinfo {author} {\bibfnamefont {J.}~\bibnamefont {Jortner}}, \
  and\ \bibinfo {author} {\bibfnamefont {U.}~\bibnamefont {Landman}},\
  }\href@noop {} {\bibfield  {journal} {\bibinfo  {journal} {Chem. Phys.
  Lett.}\ }\textbf {\bibinfo {volume} {126}},\ \bibinfo {pages} {495} (\bibinfo
  {year} {1986})}\BibitemShut {NoStop}%
\bibitem [{\citenamefont {Scheidemann}\ \emph {et~al.}(1993)\citenamefont
  {Scheidemann}, \citenamefont {Schilling},\ and\ \citenamefont
  {Toennies}}]{scheidemann1993anomalies}%
  \BibitemOpen
  \bibfield  {author} {\bibinfo {author} {\bibfnamefont {A.}~\bibnamefont
  {Scheidemann}}, \bibinfo {author} {\bibfnamefont {B.}~\bibnamefont
  {Schilling}}, \ and\ \bibinfo {author} {\bibfnamefont {J.~P.}\ \bibnamefont
  {Toennies}},\ }\href@noop {} {\bibfield  {journal} {\bibinfo  {journal} {J.
  Phys. Chem.}\ }\textbf {\bibinfo {volume} {97}},\ \bibinfo {pages} {2128}
  (\bibinfo {year} {1993})}\BibitemShut {NoStop}%
\bibitem [{\citenamefont {Closser}\ \emph {et~al.}(2014)\citenamefont
  {Closser}, \citenamefont {Gessner},\ and\ \citenamefont
  {Head-Gordon}}]{closser2014simulations}%
  \BibitemOpen
  \bibfield  {author} {\bibinfo {author} {\bibfnamefont {K.~D.}\ \bibnamefont
  {Closser}}, \bibinfo {author} {\bibfnamefont {O.}~\bibnamefont {Gessner}}, \
  and\ \bibinfo {author} {\bibfnamefont {M.}~\bibnamefont {Head-Gordon}},\
  }\href@noop {} {\bibfield  {journal} {\bibinfo  {journal} {J. Chem. Phys.}\
  }\textbf {\bibinfo {volume} {140}},\ \bibinfo {pages} {134306} (\bibinfo
  {year} {2014})}\BibitemShut {NoStop}%
\bibitem [{\citenamefont {Seong}\ \emph {et~al.}(1998)\citenamefont {Seong},
  \citenamefont {Janda}, \citenamefont {Halberstadt},\ and\ \citenamefont
  {Spiegelmann}}]{seong1998short}%
  \BibitemOpen
  \bibfield  {author} {\bibinfo {author} {\bibfnamefont {J.}~\bibnamefont
  {Seong}}, \bibinfo {author} {\bibfnamefont {K.~C.}\ \bibnamefont {Janda}},
  \bibinfo {author} {\bibfnamefont {N.}~\bibnamefont {Halberstadt}}, \ and\
  \bibinfo {author} {\bibfnamefont {F.}~\bibnamefont {Spiegelmann}},\
  }\href@noop {} {\bibfield  {journal} {\bibinfo  {journal} {J. Chem. Phys.}\
  }\textbf {\bibinfo {volume} {109}},\ \bibinfo {pages} {10873} (\bibinfo
  {year} {1998})}\BibitemShut {NoStop}%
\bibitem [{\citenamefont {Barranco}\ \emph {et~al.}(2006)\citenamefont
  {Barranco}, \citenamefont {Guardiola}, \citenamefont {Hern{\'a}ndez},
  \citenamefont {Mayol}, \citenamefont {Navarro},\ and\ \citenamefont
  {Pi}}]{Barranco:2006}%
  \BibitemOpen
  \bibfield  {author} {\bibinfo {author} {\bibfnamefont {M.}~\bibnamefont
  {Barranco}}, \bibinfo {author} {\bibfnamefont {R.}~\bibnamefont {Guardiola}},
  \bibinfo {author} {\bibfnamefont {S.}~\bibnamefont {Hern{\'a}ndez}}, \bibinfo
  {author} {\bibfnamefont {R.}~\bibnamefont {Mayol}}, \bibinfo {author}
  {\bibfnamefont {J.}~\bibnamefont {Navarro}}, \ and\ \bibinfo {author}
  {\bibfnamefont {M.}~\bibnamefont {Pi}},\ }\href@noop {} {\bibfield  {journal}
  {\bibinfo  {journal} {J. Low Temp. Phys.}\ }\textbf {\bibinfo {volume}
  {142}},\ \bibinfo {pages} {1} (\bibinfo {year} {2006})}\BibitemShut {NoStop}%
\bibitem [{\citenamefont {Hernando}\ \emph {et~al.}(2012)\citenamefont
  {Hernando}, \citenamefont {Barranco}, \citenamefont {Pi}, \citenamefont
  {Loginov}, \citenamefont {Langlet},\ and\ \citenamefont
  {Drabbels}}]{Hernando:2012}%
  \BibitemOpen
  \bibfield  {author} {\bibinfo {author} {\bibfnamefont {A.}~\bibnamefont
  {Hernando}}, \bibinfo {author} {\bibfnamefont {M.}~\bibnamefont {Barranco}},
  \bibinfo {author} {\bibfnamefont {M.}~\bibnamefont {Pi}}, \bibinfo {author}
  {\bibfnamefont {E.}~\bibnamefont {Loginov}}, \bibinfo {author} {\bibfnamefont
  {M.}~\bibnamefont {Langlet}}, \ and\ \bibinfo {author} {\bibfnamefont
  {M.}~\bibnamefont {Drabbels}},\ }\href@noop {} {\bibfield  {journal}
  {\bibinfo  {journal} {Phys. Chem. Chem. Phys.}\ }\textbf {\bibinfo {volume}
  {14}},\ \bibinfo {pages} {3996} (\bibinfo {year} {2012})}\BibitemShut
  {NoStop}%
\bibitem [{\citenamefont {Buchenau}\ \emph {et~al.}(1991)\citenamefont
  {Buchenau}, \citenamefont {Toennies},\ and\ \citenamefont
  {Northby}}]{Buchenau1991}%
  \BibitemOpen
  \bibfield  {author} {\bibinfo {author} {\bibfnamefont {H.}~\bibnamefont
  {Buchenau}}, \bibinfo {author} {\bibfnamefont {J.}~\bibnamefont {Toennies}},
  \ and\ \bibinfo {author} {\bibfnamefont {J.}~\bibnamefont {Northby}},\
  }\href@noop {} {\bibfield  {journal} {\bibinfo  {journal} {J. Chem. Phys.}\
  }\textbf {\bibinfo {volume} {95}},\ \bibinfo {pages} {8134} (\bibinfo {year}
  {1991})}\BibitemShut {NoStop}%
\bibitem [{\citenamefont {von Haeften}\ \emph {et~al.}(2011)\citenamefont {von
  Haeften}, \citenamefont {Laarmann}, \citenamefont {Wabnitz}, \citenamefont
  {Möller},\ and\ \citenamefont {Fink}}]{Haeften2011}%
  \BibitemOpen
  \bibfield  {author} {\bibinfo {author} {\bibfnamefont {K.}~\bibnamefont {von
  Haeften}}, \bibinfo {author} {\bibfnamefont {T.}~\bibnamefont {Laarmann}},
  \bibinfo {author} {\bibfnamefont {H.}~\bibnamefont {Wabnitz}}, \bibinfo
  {author} {\bibfnamefont {T.}~\bibnamefont {M\"{o}ller}}, \ and\ \bibinfo
  {author} {\bibfnamefont {K.}~\bibnamefont {Fink}},\ }\href@noop {} {\bibfield
   {journal} {\bibinfo  {journal} {J. Phys. Chem. A}\ }\textbf {\bibinfo
  {volume} {115}},\ \bibinfo {pages} {7316} (\bibinfo {year}
  {2011})}\BibitemShut {NoStop}%
\bibitem [{\citenamefont {Nijjar}\ \emph {et~al.}(2018)\citenamefont {Nijjar},
  \citenamefont {Krylov}, \citenamefont {Prezhdo}, \citenamefont {Vilesov},\
  and\ \citenamefont {Wittig}}]{nijjar2018conversion}%
  \BibitemOpen
  \bibfield  {author} {\bibinfo {author} {\bibfnamefont {P.}~\bibnamefont
  {Nijjar}}, \bibinfo {author} {\bibfnamefont {A.}~\bibnamefont {Krylov}},
  \bibinfo {author} {\bibfnamefont {O.}~\bibnamefont {Prezhdo}}, \bibinfo
  {author} {\bibfnamefont {A.}~\bibnamefont {Vilesov}}, \ and\ \bibinfo
  {author} {\bibfnamefont {C.}~\bibnamefont {Wittig}},\ }\href@noop {}
  {\bibfield  {journal} {\bibinfo  {journal} {J. Phys. Chem. Lett.}\ }\textbf
  {\bibinfo {volume} {9}},\ \bibinfo {pages} {6017} (\bibinfo {year}
  {2018})}\BibitemShut {NoStop}%
\bibitem [{\citenamefont {M{\"o}ller}\ \emph {et~al.}(1999)\citenamefont
  {M{\"o}ller}, \citenamefont {von Haeften}, \citenamefont {Laarman},\ and\
  \citenamefont {von Pietrowski}}]{moller1999photochemistry}%
  \BibitemOpen
  \bibfield  {author} {\bibinfo {author} {\bibfnamefont {T.}~\bibnamefont
  {M{\"o}ller}}, \bibinfo {author} {\bibfnamefont {K.}~\bibnamefont {von
  Haeften}}, \bibinfo {author} {\bibfnamefont {T.}~\bibnamefont {Laarman}}, \
  and\ \bibinfo {author} {\bibfnamefont {R.}~\bibnamefont {von Pietrowski}},\
  }\href@noop {} {\bibfield  {journal} {\bibinfo  {journal} {Eur. Phys. J. D}\
  }\textbf {\bibinfo {volume} {9}},\ \bibinfo {pages} {5} (\bibinfo {year}
  {1999})}\BibitemShut {NoStop}%
\bibitem [{\citenamefont {Ziemkiewicz}\ \emph {et~al.}(2015)\citenamefont
  {Ziemkiewicz}, \citenamefont {Neumark},\ and\ \citenamefont
  {Gessner}}]{Ziemkiewicz:2015}%
  \BibitemOpen
  \bibfield  {author} {\bibinfo {author} {\bibfnamefont {M.~P.}\ \bibnamefont
  {Ziemkiewicz}}, \bibinfo {author} {\bibfnamefont {D.~M.}\ \bibnamefont
  {Neumark}}, \ and\ \bibinfo {author} {\bibfnamefont {O.}~\bibnamefont
  {Gessner}},\ }\href@noop {} {\bibfield  {journal} {\bibinfo  {journal} {Int.
  Rev. Phys. Chem.}\ }\textbf {\bibinfo {volume} {34}},\ \bibinfo {pages} {239}
  (\bibinfo {year} {2015})}\BibitemShut {NoStop}%
\end{thebibliography}%

\end{document}